\documentclass[reprint,
	10pt, notitlepage,
    floats, floatfix,
	amsmath, amssymb, amsfonts, 
	superscriptaddress,
	showpacs, showkeys, aps,
    nofootinbib,prd, 
]{revtex4-2}
\usepackage{pifont}
\newcommand{\xmark}{\ding{55}}%
\usepackage{booktabs}
\usepackage{amsmath, amssymb, amsfonts}
\usepackage[utf8]{inputenc} 
\usepackage{appendix}
\usepackage[dvipsnames]{xcolor}
\xdefinecolor{mylinkcolor}{rgb}{0,0,0.5}
\usepackage[bookmarksnumbered, bookmarksopen, bookmarksopenlevel=2,breaklinks=true,colorlinks=true, filecolor=mylinkcolor,citecolor=teal,linkcolor=teal,urlcolor=teal,menucolor=mylinkcolor]{hyperref}
\usepackage{cancel, aasmacros, graphicx, dcolumn, bm, lipsum, xspace, array, verbatim, color, lineno,microtype,}
\usepackage[normalem]{ulem}
\usepackage[inline]{enumitem}
\usepackage{orcidlink}
\usepackage[capitalise]{cleveref}

\Crefname{equation}{Eq.}{Eqs.}
\Crefname{figure}{Fig.}{Figs.}
\Crefname{tabular}{Tab.}{Tabs.}
\Crefname{section}{Sec.}{Secs.}
\Crefname{subsection}{Sec.}{Secs.}

\usepackage{multirow}
\usepackage{tikz}
\usetikzlibrary{shadows,shapes.geometric,arrows.meta, positioning, calc, decorations.pathmorphing}
\usetikzlibrary{}
\usepackage{tcolorbox}
\tcbuselibrary{skins}
\tikzstyle{arrow} = [thick,->,>=stealth]
\tikzstyle{line} = [thick,-]

\newcommand{\Ins}{\affiliation{Dipartimento di Scienza e Alta Tecnologia, Università dell’Insubria, via Valleggio 11, I-22100 Como, Italy}}
\newcommand{\Bic}{\affiliation{Dipartimento di Fisica “G. Occhialini”, Università degli Studi di Milano-Bicocca, Piazza della Scienza 3, 20126 Milano, Italy}}
\newcommand{\Infn}{\affiliation{INFN, Sezione di Milano-Bicocca, Piazza della Scienza 3, 20126 Milano, Italy}}
\newcommand{\bham}{\affiliation{Institute for Gravitational Wave Astronomy \& School of Physics and
Astronomy, University of Birmingham, Birmingham, B15 2TT, UK}}
\newcommand{\inaf}{\affiliation{INAF – Osservatorio Astronomico di Brera, via E. Bianchi
46, 23807, Merate, Italy}}

\begin{document}
\preprint{APS/123-QED}
\title{
The first year of LISA Galactic foreground}

\author{Riccardo~Buscicchio\orcidlink{0000-0002-7387-6754}} \email{riccardo.buscicchio@unimib.it} \Bic \Infn \bham  
\author{Federico~Pozzoli\orcidlink{0009-0009-6265-584X}}
\email{fpozzoli@unisubria.it} \Ins \Infn
\author{Daniele Chirico\orcidlink{0000-0003-4844-1588}} \Bic
\author{Alberto Sesana\orcidlink{0000-0003-4961-1606}} \Bic \Infn \inaf

\begin{abstract}
Galactic white-dwarf binaries play a central role in the inference model for the Laser Interferometer Space Antenna. In this manuscript, we employ the \texttt{bahamas} codebase to characterize, in a global-fit fashion, the reconstruction of the Galactic foreground during the first year of observation. 
To account for its statistical properties, we represent the data in time--frequency domain, and characterize the effectiveness of multiple approaches, e.g. statistically viable likelihoods, sampling schemes, segmentation widths, and gaps density. 
Our analysis yields consistent results across, with overwhelming evidence in favor of a non-stationary model in less than a month of data.
Moreover, we show robustness against the presence of additional extragalactic foregrounds, and test the suitability of our approximations on the more complex simulated data in the \emph{Yorsh} data challenge.

\end{abstract}

\maketitle
\section{Introduction}
\label{sec:introduction}
The Laser Interferometer Space Antenna (LISA)~\cite{2017arXiv170200786A} is a pioneering space-based observatory designed to detect gravitational waves (GWs) in the millihertz frequency band. Over its mission lifetime, LISA is expected to observe a wide variety of astrophysical systems, ranging from mergers of massive black hole binaries (MBHBs) across the cosmic history to inspirals of Galactic, stellar-origin compact objects~\cite{Colpi:2024}. 

LISA will resolve thousands nearly monochromatic double white dwarfs (DWDs), offering insights into the Galactic population of compact binaries. 
The LISA datastream will be signal-dominated, with numerous GW sources persistently emitting in its frequency band. This motivates the development of so-called \textit{global fits} \cite{2023PhRvD.107f3004L,2025PhRvD.111b4060K,2025PhRvD.111j3014D,2024PhRvD.110b4005S}.

The majority of Galactic WD binaries will remain unresolved due to source confusion, hence piling up in an incoherent Galactic foreground (GF)~\cite{2009CQGra..26i4030N}, dominant over the instrumental noise at frequencies $\approx 0.5$--$3 \mathrm{mHz}$.

Recent studies have highlighted expectations for an extragalactic foreground (EF)~\cite{2025arXiv251018695P,2025arXiv250618390B,2024A&A...691A.261H,2024A&A...683A.139S}, generated by the (entirely unresolved) extra-Galactic DWD population.
By contrast to the EF, the GF originates primarily from sources concentrated toward the Galactic center in a strongly anisotropic distribution. 
Therefore, due to LISA orbital motion around the Sun, a time-dependent, quasi-periodic amplitude modulation ---often referred to as \emph{cyclostationarity}--- is induced on the observed stochastic signal.
Modelling and inferring on it (i) adds discriminating power with respect to the instrumental noise, and (ii) provides inference capability on the MW structure and morphology, where electromagnetic observations are dominated by dust extinction.
In addition, non-Gaussianity arising from the Poisson nature of the GF at high frequencies~\cite{2025EPJC...85..887B} and spurious global fit residuals~\cite{2024arXiv241017180R}, may allow for additional discriminating power as shown with heavy-tailed likelihood models in literature~\cite{2025PhRvD.111b2005K}.

In previous work~\cite{2025PhRvD.111f3005P}, we proposed a full frequency-domain approach to account for cyclostationarities. We did so by modeling off-diagonal terms in the stochastic process covariance matrix. 
We applied the method to study the detectability of backgrounds from MW satellites~\cite{2024MNRAS.531.2642R}.
Our study highlighted a key limitation: at least close-to-one year of {LISA} data had to be available to reach sufficient frequency resolution to infer on the target spectral cross-correlations. 
Global-fit analyses and transients detection in low latency (e.g. massive black-hole binaries) may not be compatible with such requirement: rapid noise inference on time-segmented data, potentially accomodating for the presence of gaps, are essential.

In this work, we extend our analysis to the time–frequency domain, representing data segments as short-time Fourier transforms (STFTs). 
Approaches have been already proposed in literature, suited to a time–frequency GF representation. Some of them rely on wavelets decomposition~\cite{2022ApJ...940...10D}, some other focus on astrophysically-motivated templates~\cite{2025PhRvD.111b3025C}, pixelation~\cite{2025JCAP...04..052H}, or spherical harmonic decomposition~\cite{2025arXiv250820308C,2021MNRAS.507.5451B} to describe sources anisotropy and the induced GF modulation.
We instead turn our attention to quantifying the evidence in favour of a cyclostationary GF during the first year of LISA operations, and to assess the impact of data segmentation, scheduled and unscheduled gaps, likelihood models, stochastic samplers, evidence estimators, and the presence of an EF component.
In addition, we test our approximations against a more realistic dataset, the \emph{Yorsh} data challenge~\cite{LDC1b}, containing a full simulation of LISA instrumental noise and a Galactic population of DWDs.
Individual studies are performed through a series of inferences with the publicly available code \texttt{bahamas}~\cite{2025arXiv250622542P}, adapting the GF modulation described in~\cite{2025EPJC...85..887B}. 

The paper is organized as follows. In~\Cref{subsec:lisa_conventions}, we first introduce the basic conventions adopted in our analysis, then describe the time-frequency representation, as implemented in \texttt{bahamas},in~\Cref{subsec:lik_time_freq}. 
In~\Cref{sec:results} we present our results. 
First, in~\Cref{subsec:gf_first_year} we consider a simulated GF in the presence of instrumental noise, only. Then, we quantify the impact of data gaps and of an additional EF in~\Cref{subsec:gaps,subsec:ef}, respectively. Finally, we apply our methodology to \emph{Yorsh} in~\Cref{subsec:yorsh}. We conclude and outline future prospects in~\Cref{sec:conclusion}.

\section{Data model}\label{sec:data_model}
\subsection{Conventions for LISA \label{subsec:lisa_conventions}}
We perform our analysis assuming time-delay interferometric (TDI) data. 
TDI is a post-processing technique that suppresses laser frequency noise by combining delayed single-link measurements from LISA~\cite{Tinto:2005}. 
We adopt the noise-uncorrelated A,E, and T channels and for simplicity we assume analytical, Keplerian orbits yielding equal, constant light-travel times between spacecrafts.
Consequently, their motion is fully determined by the initial orbital parameters: 
$\alpha_0$, which sets the initial phase of the LISA barycenter, and 
$\beta_0$, defining the initial rotation in the constellation plane (i.e., the orientation of the triangular configuration in the ecliptic at $t=0$).
In the following analyses, we set $\alpha_0, \beta_0$ to zero.
We use first-generation TDIs hence safely neglect cross-correlation given the assumed orbits.
Introducing additional complexity in the LISA constellation dynamics, such as armlength breathing or unequal link noises, introduces further non-stationarities that have not yet been fully explored in literature, although are expected to be subdominant with respect to the GF~\cite{2025JCAP...06..030K}.

We model the GF envelope following~\cite{2025EPJC...85..887B,2025PhRvD.111f3005P},
and compute it under the low-frequency approximation by averaging the LISA response over the MW sky distribution, modelled as a two-dimensional Gaussian. 
Our approach yields an analytical model extremely fast to evaluate, thus allowing for a large number of inferences.  

\subsection{STFT representation\label{subsec:lik_time_freq}}

To capture the time-varying frequency content of the cyclostationary {GF} signal we divide the full data stream into $N_{\mathrm{chunk}}$ non-overlapping segments and perform an {STFT} on each. The time-frequency representation let us define a likelihood that accounts for the signal and noise stationarity, locally in each chunk.

The total log-likelihood is expressed as the sum of the log-likelihoods for each chunk:
\begin{align}
    \log \mathcal{L}(\boldsymbol{{d}} | \theta) =
    \sum_{c = 1}^{N_\mathrm{c}} \log \mathcal{L}_{W,G}^{\mathrm{chunk}}(\boldsymbol{{d}}^c | \theta),
    \label{eq:lik}
\end{align}
where $\boldsymbol{d}^c$ denotes the data segment corresponding to the $c-$th chunk, and ${\cal L}_{W,G}^{\rm chunk}$ represent the two likelihood models considered for each single chunk, conditioned on the full set of model parameters $\theta$, and described below.
Each chunk has a duration $T$ and is sampled with a cadence $dt$. Therefore, in frequency domain $N = T/dt$ points per chunk are simulated, with a sampling frequency $f_s = 1/dt$. For a given frequency range $[f_{\rm min}, f_{\rm max}]$, we define $n_f$ as the total number of frequency points falling within it. 
We adopt the following index notation:

\begin{itemize}[itemsep=3pt,topsep=0pt,parsep=2pt,partopsep=5pt]
\item[($c$)] --- time chunk, $c = 1, \dots, N_\mathrm{c}$
\item[($j$)] --- TDI channel, $j \in \{A, E\}$
\item[($k$)] --- frequency bin, $k = 1, \dots, n_f$
\item[($l$)] --- time sample within a chunk, $l = 1, \dots, N$
\end{itemize}

In what follows, we will consider equal-duration chunks. 
However, our approach can be readily applied to unequal chunk-lengths, allowing to allocate time or frequency resolution where needed.\\

\paragraph*{\textbf{Full-resolution data.}} 
First, we consider each chunk at full frequency resolution, i.e. through a discrete Fourier transform of time-domain data, defined as
\begin{align}
\label{eq:discrete_FT}
\tilde{d}^c_{jk} = \sqrt{\frac{2}{Wf_s}}\sum_{l=1}^N w_l \, d^c_{jl} \, \exp{\left[-\frac{ 2\pi \imath}{N} k\,l\right]},
\end{align}
where $w_l$ is a window function. The factor $W = \sum_{l=1}^{N} |w_l|^2$ accounts for the window normalization. In this work, we generate synthetic data directly in the frequency domain, so windowing effects do not arise during preprocessing. For normalization consistency, we choose $W = N$.
However, in~\Cref{subsec:yorsh} we use time-domain, simulated data, hence we apply the Kaiser window with shape parameter equal to 30.
Assuming perfect Gaussianity, the single-chunk \emph{Whittle} log-likelihood reads
\begin{align}
    \label{eq:whittle}
    &\log \mathcal{L}^{\mathrm{chunk}}_W (\boldsymbol{\tilde{d}}^c | \theta) =
    \nonumber \\
    &=- \frac{1}{2}\!\!\sum_{j \in \rm {\{A,E}\}}\sum_{k = 1}^{n_f} \left(\frac{|\tilde{d}^c_{jk}|^2}{S^c_{jk}(\theta)} + \log S^c_{jk}(\theta) + \log 2 \pi\right).
\end{align}
Here, the factor $2/(N f_s)$, converting the discrete Fourier amplitudes into a one-sided power spectral density (PSD) with units of Hz\(^{-1}\)~\cite{1161901}, is implicitly accounted for in the definition of \( \tilde{d}_{jk}^c \) from~\Cref{eq:discrete_FT}.\\

\paragraph*{\textbf{Averaged periodograms.}} Alternatively, to reduce data volume and likelihood computation time, we consider data preprocessed into averaged periodograms 
\begin{align}
    \hat{P}^c_{jm} = \frac{1}{n_{b_m}} \sum_{k = 1}^{n_{b_m}}|\tilde{d}_{jk}^c|^2.
\end{align}

Here, the initial $n_f$ frequency values within each chunk $c$ are compressed into $n_g$ coarse-grained points, indexed by $m = 1, \dots, n_g$ according to the following data compression scheme (where all frequencies are expressed in Hz):
\begin{align}
    \Delta &= \frac{\log_{10}(f_{\rm{max}}) - \log_{10}(f_{\rm{min}})}{n_g}\, ,\\
    \log_{10} f_m &= {\log_{10}f_{\rm{min}} + m\Delta}\, , \\
    \mathcal{F}_k &= \{m| f_{m} < f_k < f_{m+1}\} \, ,\\
    n_{b_m} & = \left|\mathcal{F}_m\right| \, .
\end{align}
Here, $\Delta$ denotes the logarithmic-interval width, $\mathcal{F}_m$ is the set of frequencies falling within the $m$-th interval, and $n_{b_m}$ its cardinality. 
In our analyses we choose $f_{\rm min} = 0.1~\mathrm{mHz}$, $f_{\rm max} = 2.9~\mathrm{mHz}$, and $n_g = 1000$.

Being each $\hat{P}^c_{jm}$ the sum of uncorrelated, squared, circular Gaussian complex random variables, it follows a \emph{Gamma} distribution. 
Accordingly, the single-chunk likelihood reads
\begin{align}
    \label{eq:gamma}
   & \log \mathcal{L}_G^{\mathrm{chunk}} (\boldsymbol{d} = \boldsymbol{\hat{P}}^c| \theta) = \nonumber\\
 &-\sum_{j \in {\rm{\{A,E\}}}}\sum_{m = 1}^{n_g}\left[\log \Gamma\left(\frac{n_{b_m}}{2}\right) +  \frac{n_{b_m}}{2} \log S^c_{jm}(\theta)\right. +\nonumber\\
&\left. - \left(\frac{n_{b_{m}}}{2} -1\right)\log \hat{P}^c_{jm} + \frac{n_{b_m}}{2}\frac{\hat{P}^c_{jm}}{S^c_{jm}(\theta)}\right].
\end{align}\\

\paragraph*{\textbf{Power spectral densities.}} We model the {PSD} in each chunk as the following sum of three components
\begin{align}
\label{eq:time_freq_spec}
    S^c_{jk}(\theta) &= \overline{M}^c_{j}(\theta_{\rm GF_{II}}) S_{k}^{\mathrm{GF}}(\theta_{\rm GF_{I}}) + \nonumber \\
    &+ S_{jk}^{(n)}({\theta_n}) + R_{jk} S_k^{\mathrm{EF}}({\theta_{\rm EF}}).
\end{align}

We adopt the phenomenological spectral model for the Galactic foreground $S_k^{\rm GF}({\theta_{\rm GF_I}})$ proposed in Ref.~\cite{Karnesis:2021}:
  \begin{align}
        \label{eq:sgwb_galactic_foreground}
        S^{\rm{GF}}(f) &= \frac{A}{2} f^{-7/3} \exp\left[-\left(\frac{f}{f_1}\right)^{\alpha}\right] \times \nonumber\\
        &\times\left[1 + \tanh\left(\frac{f_{\mathrm{knee}} - f}{f_2}\right)\right],
    \end{align}
    where $A$ is an overall amplitude scaling factor, and the spectral-tilt frequencies $f_1, f_{\mathrm{knee}}$ are functions of the observation time $T_{\rm obs}$ through the parametrizations
    \begin{align}
    \label{eq:f1}
        \log_{10} f_1 &= a_1 \log_{10}(T_{\rm obs}) + b_1, \\
    \label{eq:fknee}
        \log_{10} f_{\mathrm{knee}} &= a_{\mathrm{knee}} \log_{10}(T_{\rm obs}) + b_{\mathrm{knee}},
    \end{align}
    with $a_i$ and $b_i$ being calibrated model parameters.
We simulate data assuming the following parameters: $\log_{10}A = -43.9$, $\alpha = 1.8$, $a_1 = -0.25$, $b_1 = -2.7$, $a_{\rm knee} = -0.27$, $b_{\rm knee} = -2.47$, $\log_{10}f_2=-3.5$.

The factor $\overline{M}^c_{j}$ describes the square of the $j$-th TDI variable modulation function $M_j(t)$, averaged over the $c$-th chunk
\begin{align}
    \overline{M}^c_{j} = \frac{1}{t^c_E - t^c_S}\int_{t^c_S}^{t^c_E} {\rm d}t M^2_j(t|{\theta_{\mathrm{GF_{II}}}}), \label{eq:mbar}
\end{align}
where $t^c_S$ ($t^c_E$) denotes the start (end) time of each chunk. 
We distinguish between ${\theta_{\mathrm{GF_{I}}}}$ and ${\theta_{\mathrm{GF_{II}}}}$ to emphasize that the first describe the signal spectrum ${\theta_{\mathrm{GF_{I}}}} = \{\mathcal{A}=\log_{10}A, \log_{10}f_{\rm knee}, \log_{10}f_1,\log_{10}f_2\}$, while the latter includes all parameters affecting the time-dependent LISA modulation. 
For a bivariate Gaussian source distribution over the sky, the parameter space is defined by 
${\theta_{\mathrm{GF_{II}}}} = \{\sin \beta, \lambda, \sin \psi, \sigma_1^2, \sigma_2^2\}$.
A detailed discussion of such parametrization is provided in~\cite{2025PhRvD.111f3005P}.
Here, $\sin \beta$ and $\lambda$ describe the position of the Galactic center in Ecliptic coordinates, whose principal axes are rotated by an angle $\psi$ with respect to the Ecliptic latitude direction, and yield variances $\sigma_{1,2}^2$.
In what follows, we inject a signal consistent with the following values: $\sin \beta = -0.096$, $\lambda = -1.62 {\rm rad}$, $\sin \psi = -0.83$, $\sigma_1^2 = 0.04 {\rm rad}^2$, $\sigma_2^2 = 0.14 {\rm rad}^2$.

The last term in~\Cref{eq:time_freq_spec} describes the SGWB of extragalactic origin considered in~\Cref{subsec:ef}. 
We model it as a stationary and isotropic stochastic signal, hence we apply the chunk-independent response $R_{jk}$~\cite{2023JCAP...04..066B,2024PhRvD.109h3029P}
to obtain the observable signal predicted in LISA.
We adopt, for simulation and inference, the template spectral model from~\cite{2024A&A...691A.261H,2024A&A...683A.139S}, consistent with a broken power law with an exponential cut-off at the highest frequencies, i.e.
\begin{align}
    \label{eq:extra_foreground}
    S^{\rm EF}(f) &= A_{\rm{EF}} \left(\frac{f}{7.2~\mathrm{mHz}}\right)^{\gamma_1} 
    \left[1 + \left(\frac{f}{7.2~\mathrm{mHz}}\right)^{4.15}\right]^{\gamma_2} \times\nonumber\\
    &\times \exp\left[-\left(\frac{f}{40.2\rm{mHz}}\right)^3\right].
\end{align}

In this work, we choose to infer on the parameters ${\theta_{\rm EF}} = \{\mathcal{A}_{\rm EF} = \log_{10} A_{\rm EF}, \gamma_1, \gamma_2\}$ and we simulate data assuming ${\cal A}_{\rm EF} = -10.76$, $\gamma_1 = 0.741$, $\gamma_2 = -0.255$.
We emphasize that the cutoff observed in the EF component arises purely from physical considerations, rather than from DWD subtraction, by contrast with the GF case. 
Therefore, the spectral parameters are not a function of $T_{\rm obs}$.

\begin{figure*}[t]
    \centering
    \includegraphics[width=2\columnwidth]{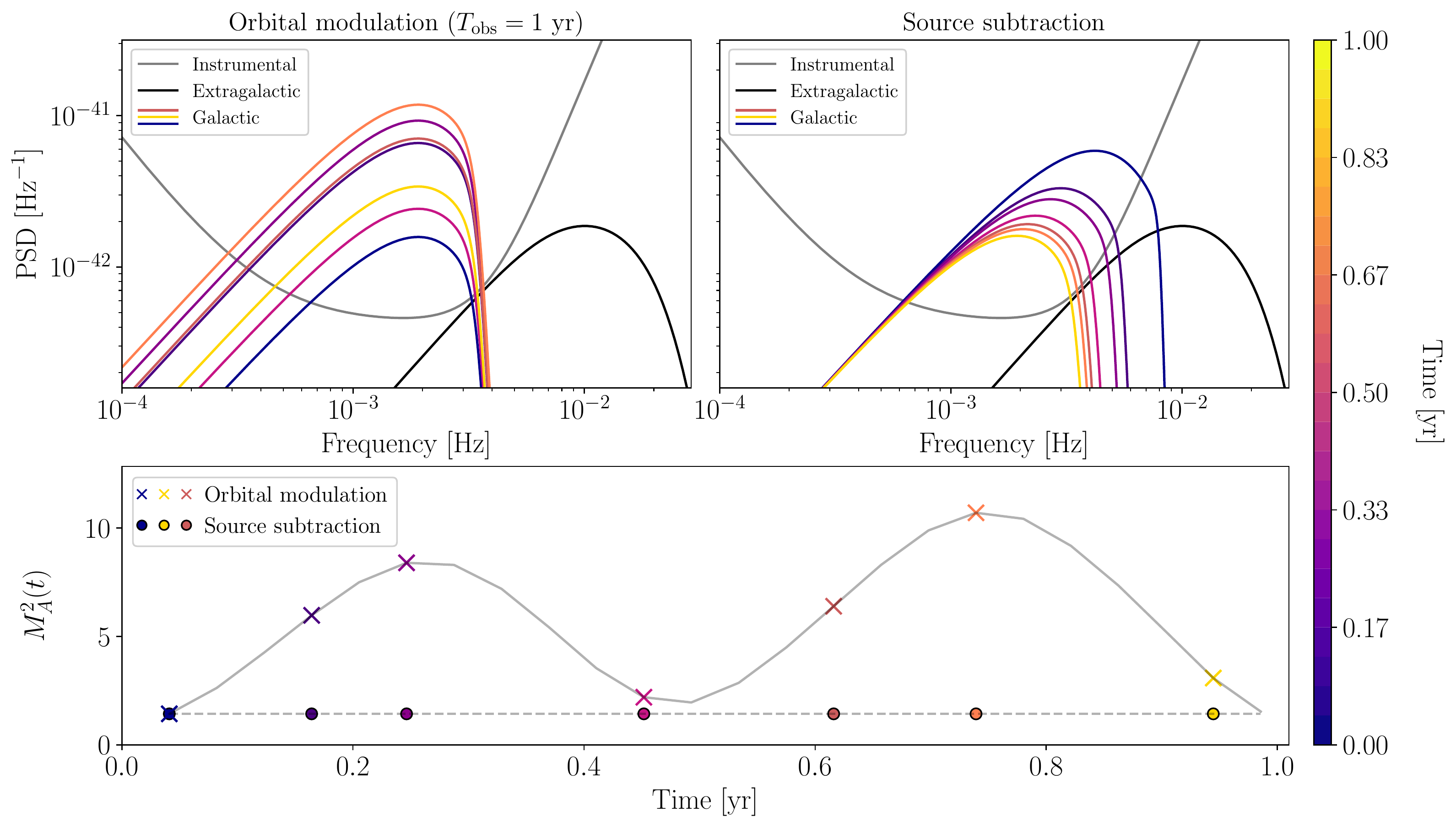}
    \caption{(\textit{Top left}) Evolution of the Galactic foreground spectrum across different time chunks, with the corresponding modulation amplitudes shown in the bottom subplot as colored crosses. The spectrum level is obtained for a fixed total observation time $T_{\rm obs}=1 {\rm yr}$. (\textit{Top right}) Evolution of the spectrum as the observation time increases. By contrast to the left panel, the modulation amplitude is fixed to that of the first chunk, shown in the bottom subplot as colored dark blue circle. The overall effect on the spectrum is a shift toward lower frequencies.
    (\textit{Bottom}) The modulation squared amplitude, highlighting the corresponding reference times as cross and circle markers.
    For simplicity, all quantities refer to the A channel.}
    \label{fig:galactic_spec}
\end{figure*}

Finally, the assumed instrumental noise model captures the two main components remaining after TDI post-processing: the optical metrology system (OMS) noise and the test mass (TM) noise. Both components are projected into the TDI channels via transfer functions, as implemented in~\cite{Nam:2023}. The PSDs of the two components read
\begin{align}
    S^n_{\rm{tm}}(f) &= \left(A \times 10^{-15} {\rm ms}^{-2} {\rm Hz}^{-1/2}\right)^2 \frac{1}{(2\pi c f)^{2}} \times \nonumber \\
    &\times \left[1 + \left(\frac{0.4\rm{mHz}}{f}\right)^2\right]\!\!\left[1 + \left(\frac{f}{8\rm{mHz}}\right)^4\right], \\
    S^n_{\rm{oms}}(f) &= \left(P \times 10^{-12} {\rm m \, Hz}^{-1/2}\right)^2 \left(\frac{2\pi f}{c}\right)^2 \times \nonumber \\
    & \times \left[1 + \left(\frac{2\rm{mHz}}{f}\right)^4\right],
\end{align}
where $c$ denotes the speed of light.
Throughout this work, we choose fixed $A = 2.4$ and $P = 7.9$ for all data segments. 

In a global fit scheme, Galactic foreground estimates have to be continuously updated as the individual source content evolves during Gibbs-sampling steps---sources are added, removed, or their parameters changed--- and as more data are accumulated during the mission.
In the absence of a full global fit pipeline, we use an iterative source subtraction scheme to approximate the {GF} evolution, following~\cite{Karnesis:2021}. 
Therefore, we adopt the following procedure:
we generate random Gaussian samples consistent with the PSD model defined in~\Cref{eq:sgwb_galactic_foreground}; then, we sequentially increase at each iteration the number of time segments, adjusting new segments spectra for their corresponding observation time, as described by~\Cref{eq:f1,eq:fknee}.

We emphasize once again that the GF PSD level in each segment (or \emph{packet}) accounts for the corresponding modulation factor, too.
Therefore, the GF PSD evolution reflects both the gradual subtraction of simulated individual DWD signals---resulting in a drift toward lower frequencies---and the modulation-induced variations, resulting in an up-and-down drift of the spectrum.

In~\Cref{fig:galactic_spec}, we illustrate the evolution of the GF spectrum across different time chunks  and as a function of the total observation time, for seven reference times across a year. 
For reference, we underplot the instrumental noise and EF spectra, both assumed stationary. In addition, we show the modulation squared amplitude, highlighting the corresponding reference times.

\paragraph*{\textbf{Code infrastructure.}}
The model described above is implemented in \texttt{bahamas}~\cite{2025arXiv250622542P}, a code flexible enough to support a variety of operational configurations.
\begin{table*}[t]
\centering
\setlength{\tabcolsep}{4pt} 
\renewcommand{\arraystretch}{1.1}
\begin{tabular}{@{}ccc@{}ccccc}
\toprule\toprule
$T$ & Analysis & Gaps & Likelihood & Sampler  & Model & Chunks no. & Ref \\
\midrule\midrule
\multirow{4}{*}{2 weeks} & \multirow{4}{*}{Sequential} & \multirow{4}{*}{\xmark} 
& \multirow{2}{*}{Whittle} & \multirow{2}{*}{HMC, NS} & Quasi-stationary (QS) & \multirow{2}{*}{1,\dots,24} & \multirow{2}{*}{\cref{fig:bayes_fac,fig:scheme,fig:corner_sum}} \\
&  & &  &  & Stationary (S) & & \\
\cmidrule(lr){4-8}
& & & \multirow{2}{*}{Gamma} & \multirow{2}{*}{HMC, NS} & QS & \multirow{2}{*}{1,\dots,24} & \multirow{2}{*}{\cref{fig:bayes_fac,fig:corner_sum,fig:corner_chunk2,fig:ridgeline}} \\
&  & &  &  & S & & \\
\midrule
\multirow{2}{*}{1 week} & \multirow{2}{*}{Sequential} & \multirow{2}{*}{scheduled} 
& \multirow{2}{*}{Gamma} & \multirow{2}{*}{NS} & QS & \multirow{2}{*}{\xmark} & \multirow{2}{*}{\cref{fig:corner_chunk2}} \\
& & &  &  & S & & \\
\midrule
\multirow{2}{*}{1 week} & \multirow{2}{*}{Sequential} & \multirow{2}{*}{sched.+unsched.} 
& \multirow{2}{*}{Gamma} & \multirow{2}{*}{NS} & QS & \multirow{2}{*}{\xmark} & \multirow{2}{*}{\cref{fig:corner_chunk2}} \\
& & &  &  & S & & \\
\midrule
2 weeks & Differential & \xmark & Gamma & NS  & QS & 1,\dots,24 & \cref{fig:evidence_envelope,fig:corner_chunk} \\
\midrule
\multirow{3}{*}{2 weeks} & \multirow{3}{*}{Sequential} & \multirow{3}{*}{\xmark} 
& \multirow{3}{*}{Gamma} & \multirow{3}{*}{NS} & QS $\cup$ EF & \multirow{3}{*}{1,\dots,24} & \multirow{3}{*}{\cref{fig:bf_ef,fig:res_mod,fig:ridgeline2}} \\
& & & & & S $\cup$ EF & & \\
& & & & & QS & & \\ 
\bottomrule\bottomrule
\end{tabular}
\caption{Summary of parameter estimations. We consider a year mission duration across all inferences. 
Columns from left to right: chunk duration $T$; analysis scheme as described in~\Cref{fig:analysis-scheme}; presence of scheduled and unscheduled gaps as detailed in~\Cref{subsec:gaps}; alternative likelihood models considered, as introduced in~\Cref{subsec:lik_time_freq}; sampling strategy adopted. Following, we list the inference model assumed, either quasi-stationary or stationary for the Galactic foreground (GF), and a stationary one for the extra-Galactic foreground (EF) and the instrumental noise --- we omit the latter from labels as the same is always adopted, as described in~\Cref{subsec:lik_time_freq}---. 
Finally, in the two rightmost columns we enumerate the data chunks considered for each analysis and provide cross-references to figures where results are presented. 
We omit listing chunks associated with analyses on gapped data, as they are better described in~\Cref{fig:corner_chunk2}.
}
\label{tab:config_summary}
\end{table*}
\begin{figure*}[htbp]
\label{fig:analysis-scheme}
\centering
\definecolor{packetcolor1}{RGB}{83,79,187}
\definecolor{packetcolor2}{RGB}{130, 51,226}
\definecolor{packetcolor3}{RGB}{130, 51,226}
\definecolor{packetcolor4}{RGB}{168,34,150}
\definecolor{packetcolor5}{RGB}{203,70,121}
\definecolor{packetcolor6}{RGB}{229,107,93}
\definecolor{packetcolor7}{RGB}{248,148,65}
\definecolor{packetcolor8}{RGB}{253,195,40}
\definecolor{packetcolor9}{RGB}{240,249,33}
\definecolor{packetcolor10}{RGB}{252,255,164}
\begin{minipage}[t]{0.48\textwidth}
\centering
\begin{tikzpicture}[scale=0.9, transform shape,
    block/.style={draw, minimum height=0.8cm, minimum width=1.4cm, align=center, fill opacity=0.7},
    prior/.style={block, fill=gray!50, fill opacity=1.0},
    noise/.style={block, fill=green!70!black, opacity=1.0},
    node distance=1cm and 0.5cm,
    every node/.style={font=\scriptsize}
]
\node[prior] (prior1) {Prior\\$\pi(\theta)$};
\node[block, fill=packetcolor1, right=of prior1] (pack1a) {Packet 1};
\node[noise, right=of pack1a] (noise1) {Posterior\\$p(\theta \mid d)$};
\draw[->] (prior1) -- (pack1a);
\draw[->] (pack1a) -- (noise1);
\node[prior, below=1.cm of prior1] (prior2) {Prior\\$\pi(\theta)$};
\node[block, fill=packetcolor1, right=of prior2] (p1b) {Packet 1};
\node[block, fill=packetcolor3, right=of p1b] (p2b) {Packet 2};
\node[noise, right=of p2b] (n2) {Posterior\\$p(\theta|d)$};
\draw[->] (prior2) -- (p1b);
\draw[->] (p1b) -- (p2b);
\draw[->] (p2b) -- (n2);
\node[prior, below=3cm of prior1] (prior3) {Prior\\$\pi(\theta)$};
\node[block, fill=packetcolor1, right=of prior3] (p1c) {Packet 1};
\node[block, fill=packetcolor3, right=of p1c] (dots) {$\cdots$};
\node[block, fill=packetcolor8, right=of dots] (pnc) {Packet $n$};
\node[noise, right=of pnc] (n3) {Posterior\\$p(\theta|d)$};
\draw[->] (prior3) -- (p1c);
\draw[->] (p1c) -- (dots);
\draw[->] (dots) -- (pnc);
\draw[->] (pnc) -- (n3);
\end{tikzpicture}
\end{minipage}
\hfill
\begin{minipage}[t]{0.48\textwidth}
\centering
\begin{tikzpicture}[scale=.9, transform shape,
    block/.style={draw, minimum height=0.8cm, minimum width=1.4cm, align=center, fill opacity=.7},
    faded/.style={block, fill=gray!10, draw=none, fill opacity=1.0},
    prior/.style={block, fill=gray!50, fill opacity=1.0},
    noise/.style={block, fill=green!70!black, fill opacity=1.0},
    every node/.style={font=\scriptsize},
    node distance=1cm and 0.5cm,
    curved arrow/.style={->, to path={.. controls +(1.0,1.0) and +(-1.0,1.0) .. (\tikztotarget)}}
]
\node[prior] (prior1b) {Prior\\$\pi(\theta)$};
\node[block, fill=packetcolor1, right=of prior1b] (pack1a) {Packet 1};
\node[noise, right=of pack1a] (noise1) {Posterior\\$p(\theta \mid d_1)$};
\draw[->] (prior1b) -- (pack1a);
\draw[->] (pack1a) -- (noise1);
\node[prior, below=1.cm of prior1b] (prior2b) {Prior\\$\pi(\theta)$};
\node[faded, right=of prior2b] (faded1) {Packet 1};
\node[block, fill=packetcolor3, right=of faded1,] (pack2b) {Packet 2};
\node[noise, right=of pack2b] (noise2) {Posterior\\$p(\theta \mid d_2)$};
\draw[curved arrow] (prior2b) to (pack2b);
\draw[->] (pack2b) -- (noise2);
\node[prior, below=3.cm of prior1b] (prior3b) {Prior\\$\pi(\theta)$};
\node[faded, right=of prior3b] (faded2) {Packet 1};
\node[faded, right=of faded2] (dots3) {$\cdots$};
\node[block, fill=packetcolor8, right=of dots3] (packn) {Packet $n$};
\node[noise, right=of packn] (noise3) {Posterior\\$p(\theta \mid d_n)$};
\draw[curved arrow] (prior3b) to (packn);
\draw[->] (packn) -- (noise3);
\end{tikzpicture}
\end{minipage}
\caption{Data acquisitions and analysis schemes considered in this work. (\textit{Left panel}) Sequential analysis scheme: data segments are generated and cumulatively analyzed over time. 
Results from this approach are shown in ~\Cref{fig:bayes_fac,fig:ridgeline,fig:ridgeline2,fig:corner_chunk2}. (\textit{Right panel}) Differential analysis scheme: 2 weeks-long data segments are analyzed independently, and the modulation is reconstructed from in-segment PSD evolution, i.e. across first and second week. Results from this approach are shown in \Cref{fig:evidence_envelope,fig:corner_chunk}.}
\label{fig:scheme}
\end{figure*}
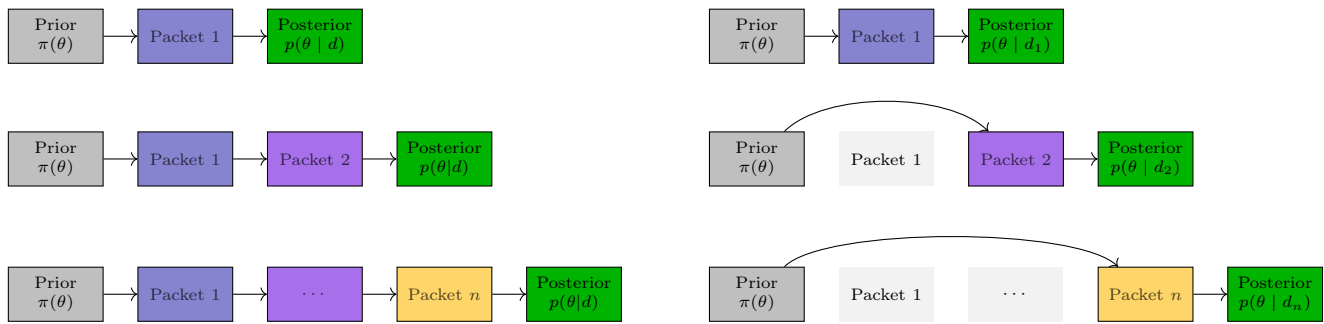
First, parameter estimation can be performed using either nested sampling (NS) or the  No-U-Turn Sampler (NUTS) variant of Hamiltonian Monte Carlo (HMC), as implemented in \texttt{nessai}~\cite{2021PhRvD.103j3006W} and \texttt{NumPyro}~\cite{2019arXiv191211554P}, respectively. 
While the former yields directly marginal-likelihood estimates, the latter is a more suitable candidate for deployment in a global-fit infrastructure. 
The typically large data volume at each likelihood evaluation makes HMC particularly appealing, as native support for automatic differentiation and accelerated hardware in \texttt{Numpyro} offsets dramatically the posterior exploration time.
Second, \texttt{bahamas} hosts implementations of both the Whittle likelihood on STFT data and the Gamma likelihood on averaged power spectra, as described in~\Cref{eq:whittle,eq:gamma}.
Third, \texttt{bahamas} can flexibly simulate and analyse segments of arbitrary, heterogeneous lengths.
For simplicity, we will consider equal-length segments of 1 and 2 weeks, both yielding frequency content well below $0.1\,\mathrm{mHz}$.

\section{Results}
\label{sec:results}
Our results are organized as follows. 
In~\Cref{subsec:gf_first_year} we first present inferences, on simulated GF and instrumental noise, only. 
We show the equivalence of the two likelihood models introduced in~\Cref{eq:whittle,eq:gamma}, and quantify the evidence in favour of a cyclostationary GF model during the first year of LISA operations.
We argue for the suitability of our approach, by showing the equivalence of posteriors obtained through nested sampling and HMC.
Then, we investigate the impact of data gaps in~\Cref{subsec:gaps}, and of an additional EF component in~\Cref{subsec:ef}. 
Finally, we apply our methodology to the more realistic \emph{Yorsh} data challenge~\cite{LDC1b} in~\Cref{subsec:yorsh}.
A summary of inferences performed is provided in \Cref{tab:config_summary}. Additional, auxiliary plots are shown in \Cref{app:plot}. 
Overall, our model parameter space consists of 14 real numbers and we adopt uniform priors over $\theta_{\rm GF_I}$, $\theta_{\rm GF_{II}}$, $\theta_{\rm EF}$, and $A,P$.

\subsection{Galactic Foreground across the first year \label{subsec:gf_first_year}}

\begin{figure*}[t]
    \centering
    \includegraphics[width=2\columnwidth]{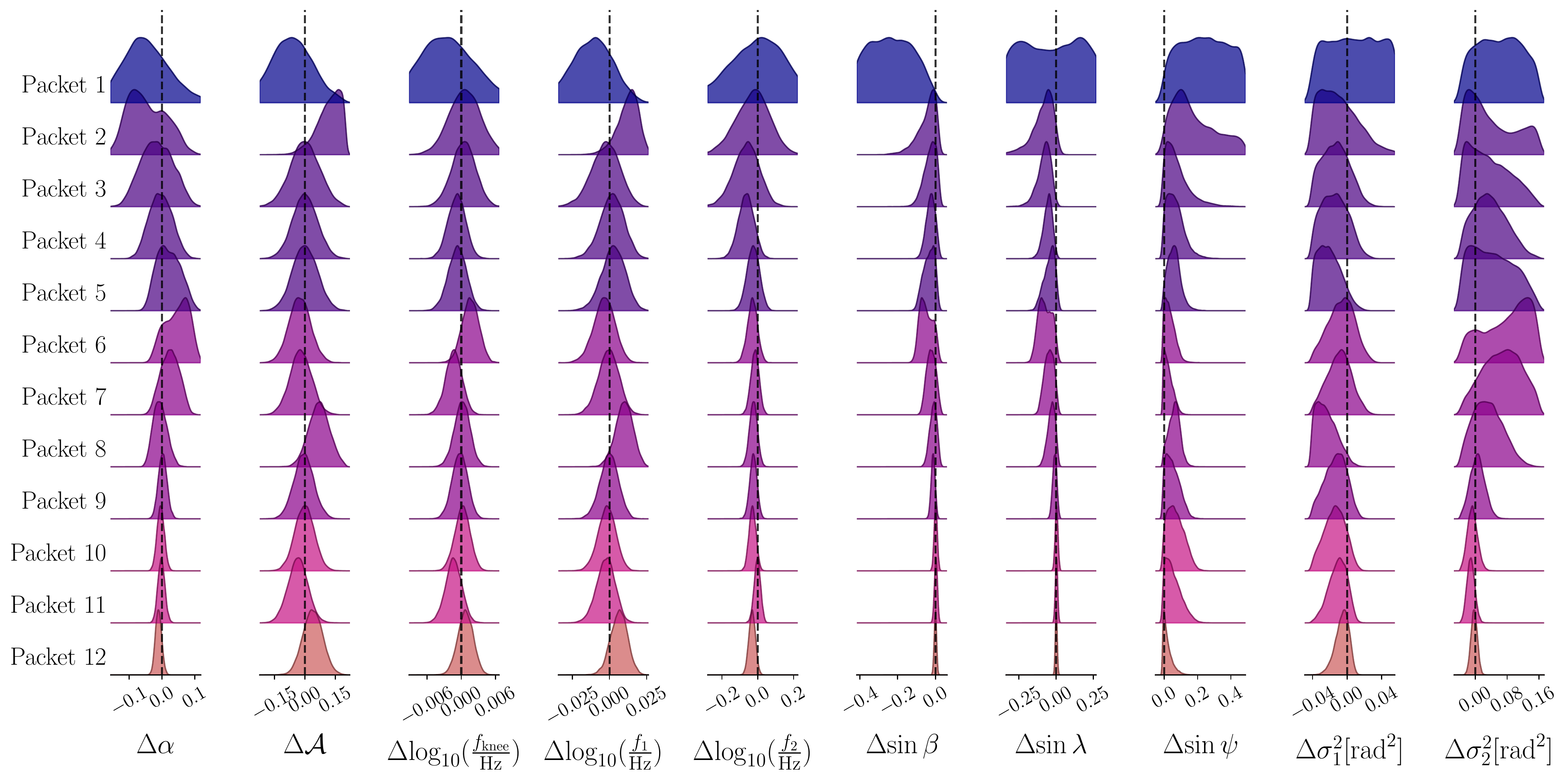}\\
    \includegraphics[width=2\columnwidth]{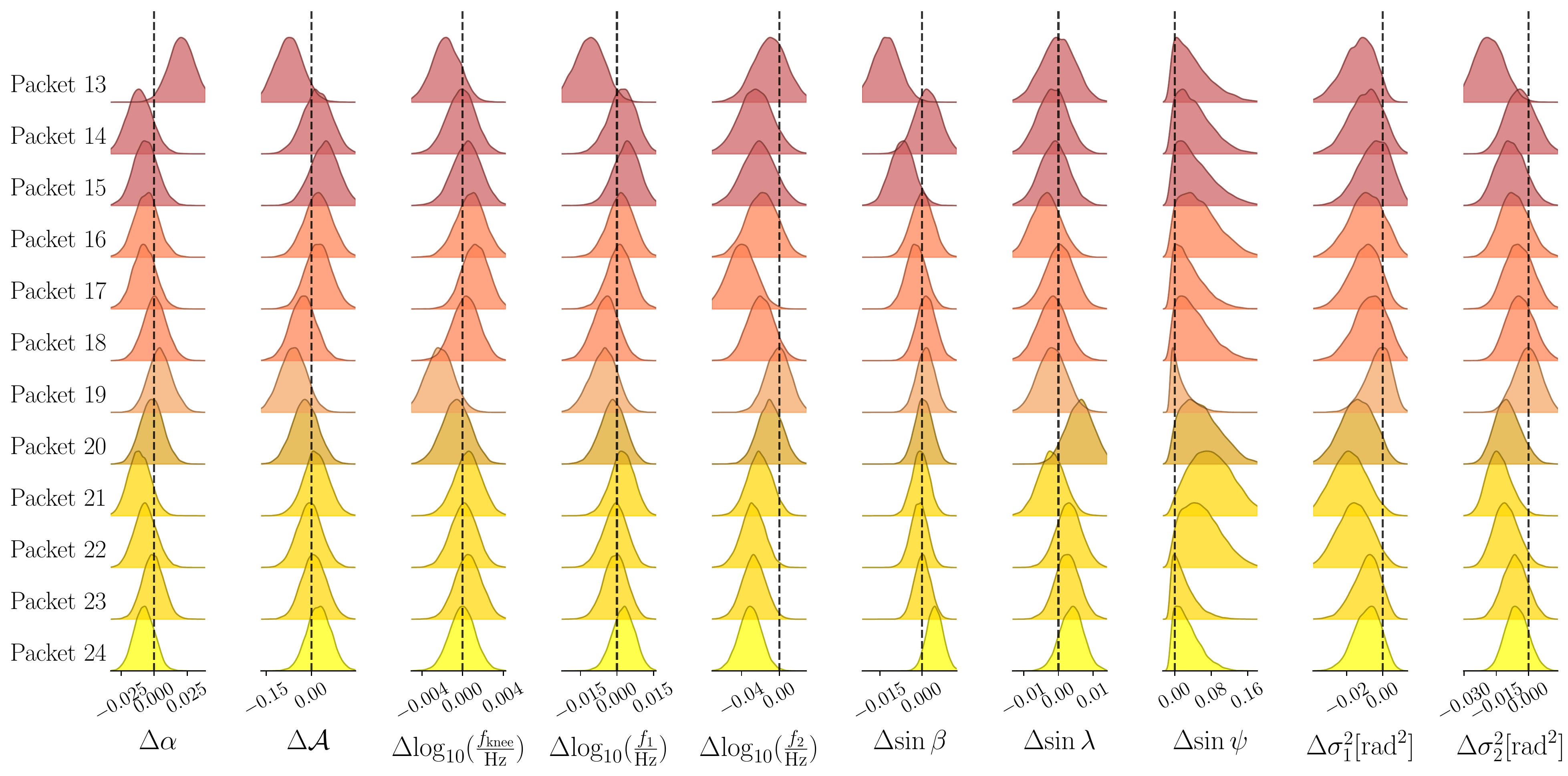}
    \caption{Ridgeline plot of the posterior distributions of Galactic foreground parameters obtained from a sequential analysis over one year of observation, using the Gamma likelihood and two-weeks-long segments.
    Posteriors are centered on the injected values, indicating unbiased reconstructions.
    From top to bottom, posterior distributions narrow down as more data are accumulated. The Galactic modulation parameters become increasingly well-constrained by leveraging consistency across different segments, up to and including the packet shown on the leftmost column. To help visualization, we scale axis ranges in the bottom panel to the typical posterior widths over the second half year.}
    \label{fig:ridgeline}
\end{figure*}

We consider a first analysis scheme mimicking a \textit{global-fit-like} approach, as illustrated in the left panel of~\Cref{fig:scheme}. As in~\Cref{tab:config_summary,fig:scheme}, we refer to it as the \emph{sequential} analysis. At each iteration we produce posterior distributions for both the instrumental noise and the GF for increasingly longer, segmented datasets.
\begin{figure}[t]
    \centering
    \includegraphics[width=1\columnwidth]{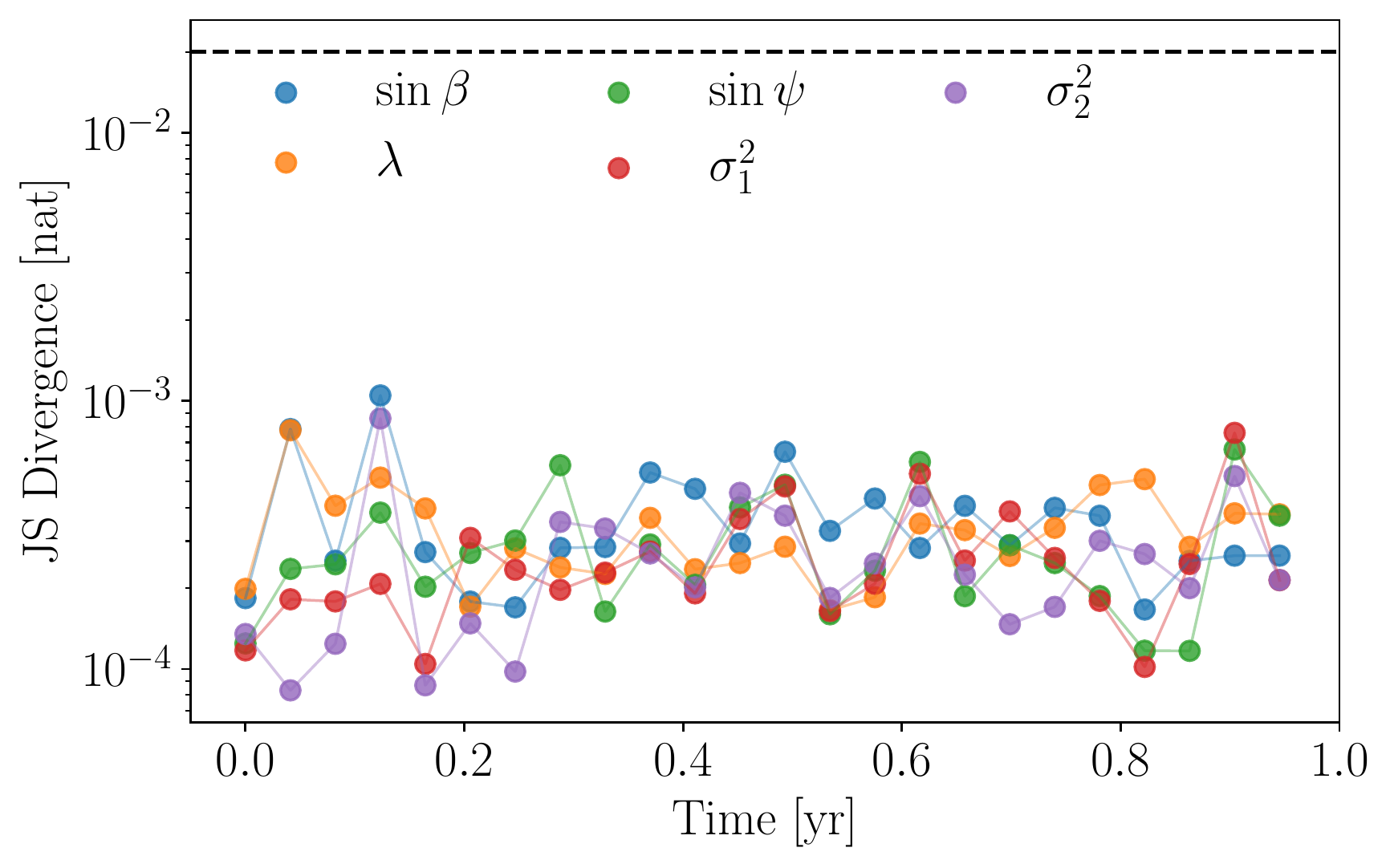}
    \caption{Jensen–Shannon divergence between the marginal posterior distributions on the modulation parameters, as inferred from the Gamma and Whittle likelihoods, and as a function of observation time.
    Each value remains well below 0.02~$\rm{nat}$ (black dashed line), a rough threshold to establish indistinguishability between one-dimensional distributions.
}
\label{fig:bayes_fac}
\end{figure}
\paragraph*{\textbf{Posterior evolution and comparison.}}
We first inject and recover the GF and LISA instrumental noise, only.
Marginal posteriors for the former at each iteration are shown in~\Cref{fig:ridgeline}, where we highlight the progressive improvement in parameter estimation as more data are accumulated.

As expected, the Whittle and Gamma likelihoods yield consistent posterior distributions~\cite{2025arXiv250524695F}.
Asymptotically, this is due to averaged periodograms being a sufficient statistics for the spectral density. Hence, thanks to the Rao-Blackwell theorem~\cite{Blackwell1947,Rao1992}, Gamma-based estimates are as unbiased as the Whittle-based ones. Moreover, variance of the former are upper-bounded by those of the latter.
To further verify our findings, we compute the Jensen–Shannon divergence (JSD) between the posterior distributions obtained from the two likelihoods. This metric quantifies the similarity between two distributions, $P$ and $Q$ \cite{61115} and it reads
\begin{align}
\mathrm{JSD}(P || Q) &= \tfrac{1}{2}D(P || M ) + \tfrac{1}{2}D(Q || M ),
\end{align}
with 
\begin{align}
D(P||M) &= \int_{X} {\rm d}x p(x) \log \left(\frac{p(x)}{m(x)}\right) ,
\end{align}
where $M = \tfrac{1}{2}(P+Q)$.
For clarity, we display in~\Cref{fig:bayes_fac} only results on divergences for ${\theta_{\rm GF_{II}}}$, for each iteration of the sequential analysis.
The observed values over time suggest that the posterior distributions derived from the two likelihoods are largely compatible.
Likewise, different stochastic samplers yield very similar, although not shown explicitly in this work.
We illustrate consistency across samplers and likelihoods in the full parameter space with a representative corner plot in~\Cref{fig:corner_sum}, obtained after 34 weeks of observation: we present three joint posterior distributions, obtained from NS and Whittle likelihood, NS with Gamma likelihood, and NUTS with Gamma likelihood, respectively.
The agreement across the different methods is largely satisfactory over the full parameter space.

By yielding the same posteriors under the same prior assumptions, both likelihoods are expected to produce consistent values of marginal-likelihood, defined as
\begin{equation}
\mathcal{Z}(d) = \int {\rm d}\theta \mathcal{L}(d|\theta)\pi(\theta)\, .
\end{equation}
While NS algorithms naturally compute the evidence as part of their inference process~\cite{10.1214/06-BA127}, Monte Carlo Markov-chain methods such as NUTS do not provide direct estimates of it. 
Nevertheless, several approaches have been developed to infer the evidence from MCMC samples, e.g. thermodynamic integration and the stepping stone (SS) algorithm~\cite{Maturana:2019}. 
We verify that a generalized version of the SS algorithm~\cite{2025MNRAS.540.3818Z} returns evidence estimates consistent with those obtained via NS, with a relative discrepancy smaller than
$1\%$ in $\log \mathcal{Z}$.

\begin{figure}[t]
    \centering
    \includegraphics[width=1\columnwidth]{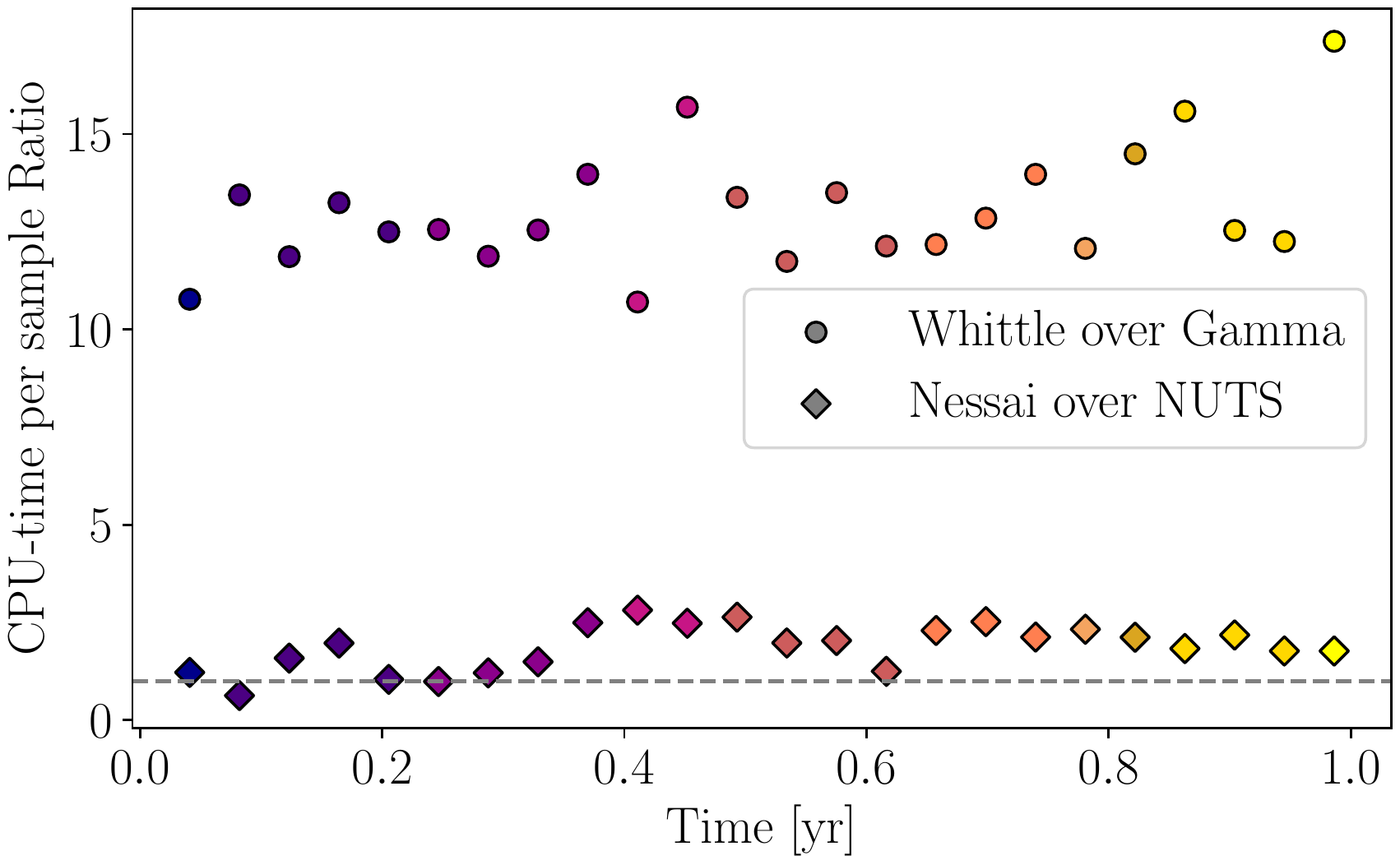}
    \caption{
        Computational gain factor across different setup configurations available in \texttt{bahamas}, as a function of data length. 
        Markers denote the ratio of CPU time required to obtain one posterior sample in two competing configurations:
        circles denote the ratio of Whittle over Gamma likelihoods, while diamonds denote the ratio of NS over NUTS samplers. 
        Each marker is colored according to the total observation time considered, matching those in \Cref{fig:ridgeline}.
        While the Gamma likelihood is approximately 15 times faster than Whittle, NUTS yields samples twice as fast as NS. The latter is furthermore expected to yield even larger speedups once deployed on GPU.
}
\label{fig:computational_gain}
\end{figure}

In~\Cref{fig:computational_gain}, we show the performance gain achieved under different setups.
Specifically, posterior sampling with NUTS---fully relying on \texttt{jax} for reverse-mode differentiation---results in a speed-up factor of approximately $2.0$.
In this work, we do not exploit GPU acceleration (natively supported by \texttt{jax}), so an even larger gain is expected in such scenario.
The use of the Gamma likelihood yields a significantly larger reduction in computational cost.
This depends on the exploitation of power spectra averaging, which allows to reduce the number of computations per-likelihood-evaluation from $n_f$ to $n_g\ll n_f$.
The adopted averaging and binning scheme introduced in ~\Cref{sec:introduction} results in additional speed-up factor of about $15.0$. Overall, a total computational gain of roughly a factor $30$ is achieved.
Based on our findings, we henceforth adopt the Gamma likelihood, as it is significantly faster to evaluate while providing unbiased results with respect to the full-frequency resolution Whittle. 

\paragraph*{\textbf{Bayes Factor time evolution}}
\begin{figure*}
    \centering
    \includegraphics[width=1\columnwidth]{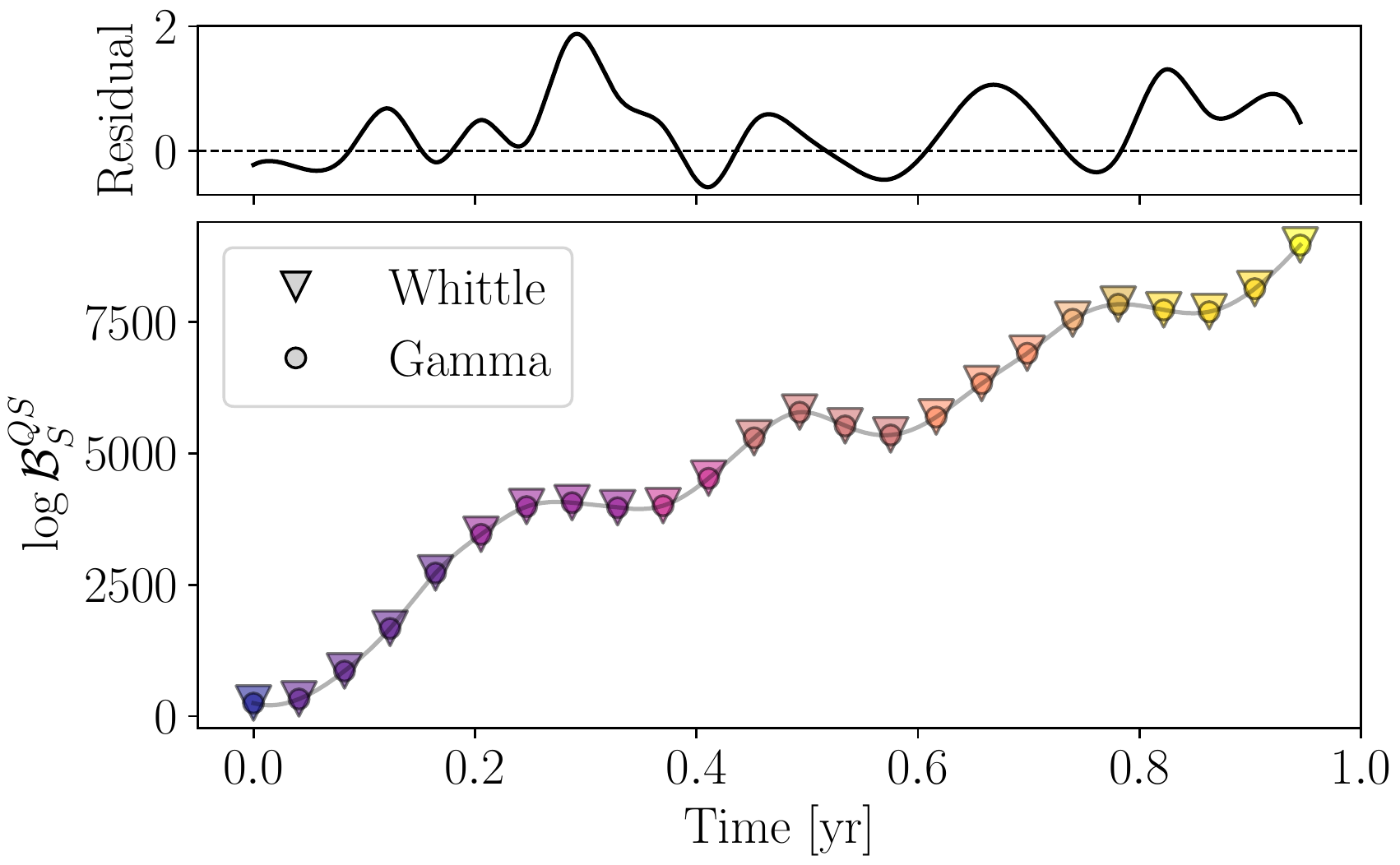}
    \includegraphics[width=1\columnwidth]{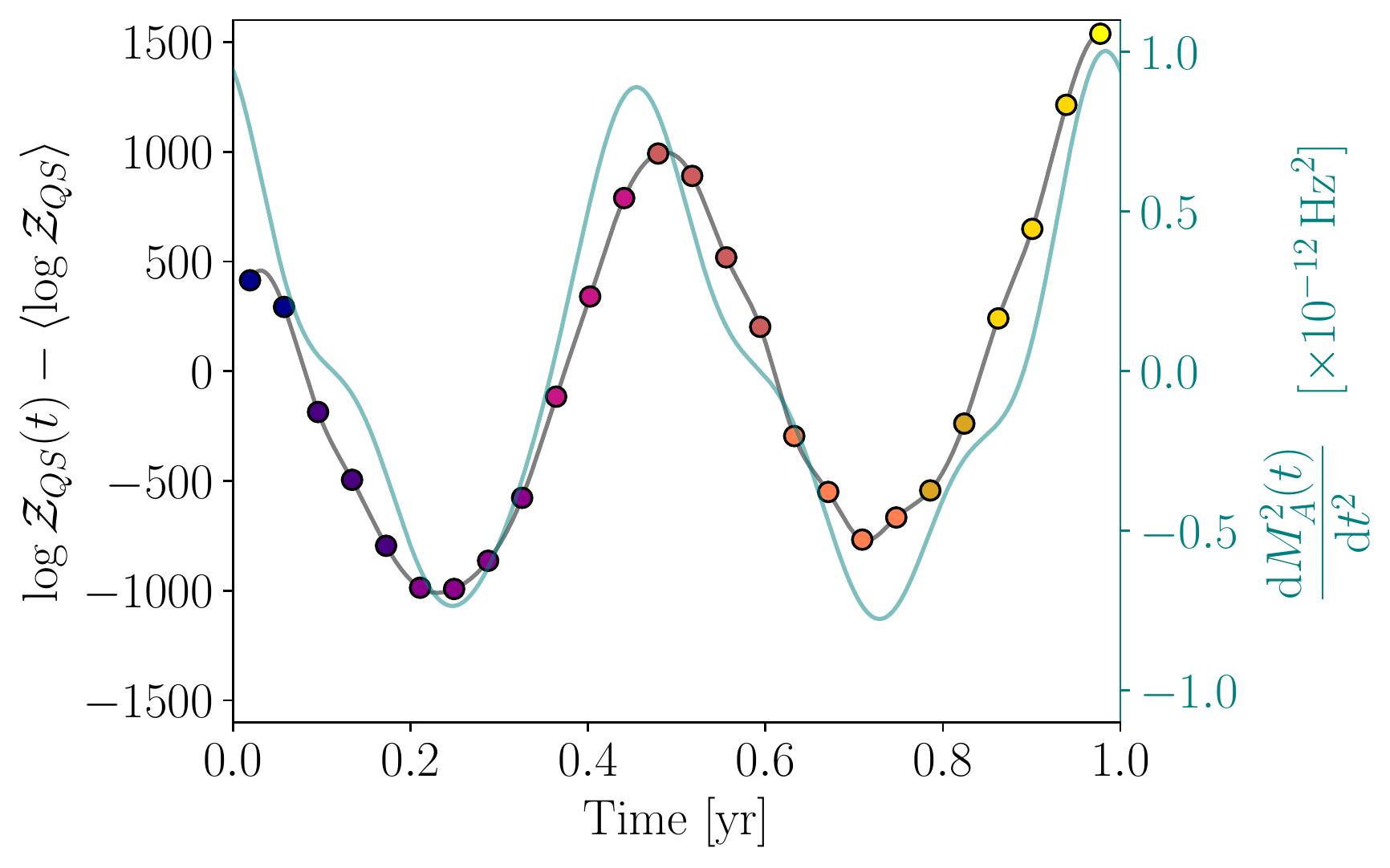}
    \caption{(\textit{Left Panel})
    Temporal evolution of the $\log$ Bayes factor between the quasi-stationary and stationary hypotheses for the Galactic foreground, obtained from a sequential analysis. The residual plot above shows the absolute difference between Bayes factors obtained with the Gamma and Whittle likelihood, respectively. (\textit{Right Panel}) Evolution of the log-evidence for the quasi-stationary model over time. Each colored circle corresponds to the log-evidence obtained from  two consecutive weeks of data. Values are shifted by the mean log-evidence across all segments considered. The teal axis refers to the second time derivative ${{\rm d}{M}^2_A(t)}/{{\rm d} t^2}$ of the A channel squared modulation, shown as teal solid line.
    }
\label{fig:evidence_envelope}
\end{figure*}
We now assess the statistical suitability of the quasi-stationary model, as compared to a simpler, stationary one.
We compute the (log-)Bayes factor, defined as the (log-)ratio of marginal likelihoods between the quasi-stationary hypothesis and the stationary one
\begin{align}
    \log \mathcal{B}_{S}^{QS} = \log \mathcal{Z}_{QS} - \log \mathcal{Z}_{S}\,,
\end{align}
where the stationary assumption corresponds to setting 
$\overline{M}^i_j = 1$ in~\Cref{eq:sgwb_galactic_foreground}.
We show in the left panel of~\Cref{fig:evidence_envelope} the evolution of $\mathcal{B}$ over time for both the Whittle and Gamma likelihoods. As anticipated in~\Cref{subsec:gf_first_year}, the two approaches are equivalent.
The evidence in favour of the quasi-stationary hypothesis increases significantly over time. After the first few weeks of data are accumulated, the quasi-stationary hypothesis is overwhelmingly favored. 
A more complex, oscillating and upwards drifting structure, emerges over longer $T_{\rm obs}$, likely tracking the shape of the Galactic modulation.

To investigate the observed trend, we perform a `differential' analysis (as referred to in~\Cref{tab:config_summary} and right panels of ~\Cref{fig:scheme,fig:evidence_envelope}), where 
the year of simulated data is divided into two weeks segments. Unlike the previous analysis, we now examine each pair of consecutive weeks as two chunks, using the quasi-stationary model.
By tracking each segment evidence over time, we aim to identify which segment yields the largest constraining power, likely driven by the varying modulation pattern across consecutive chunks.
The evolution of the model evidence over time is shown in the right panel of \Cref{fig:evidence_envelope}. Interestingly, the evidence does not track the modulation directly; rather, it appears to follow the second time derivative ${{\rm d}{M}^2_A(t)}/{{\rm d} t^2}$ of the squared modulation amplitude. 
The smooth, rapid variations in the modulation help break degeneracies between adjacent segments, thus constraining the fundamental harmonics of the modulation model~\cite{2025EPJC...85..887B}, eventually  allowing the QS model to better constrain the signal over time.
This behavior has important implications for parameter reconstruction, which we illustrate in~\Cref{fig:corner_chunk}, focusing on three distinct periods of increasing $\log$-evidence throughout the year. 
We omit the reconstruction of $\sin \psi$, $\sigma_1^2$ and $\sigma_2^2$, as they are poorly constrained when inferred from two-weeks data segments, only.
We observe that the reconstruction of the source sky distribution progressively and steadily improves. 
In particular, after 3 months, the inference is primarily driven by the spectral shape parameters, which are better constrained in the early stages as compared to $\sin \beta$ and $\lambda$. 
Conversely, the differential inference on the last segment yield tight constrains on the quasi-stationarity of the signal, while those on the spectral parameters are not significantly improved.
As expected, an intermediate behavior is observed after a few months of data.

\subsection{Impact of data gaps\label{subsec:gaps}}
We now quantify the effect of datastream gaps on the reconstruction of the Galactic foreground, as presented in~\Cref{subsec:gf_first_year}.
Periods of unavailable data due to maintainance, antenna-repointing, or unforeseen exogenous events can be broadly classified either as \textit{scheduled} or \textit{unscheduled} \textit{gaps}.

In contrast to previous studies~\cite{2025PhRvD.111l4053B}, in \texttt{bahamas} we do not explicitly model the presence of data gaps \emph{in the data}. 
Instead, assuming that the start and end times of each gap are known, we exclude the corresponding data from our inference. To explore their effect on our results, we simulate both types of gaps, following the approach in~\cite{2021PhRvD.104d4035D}:
\begin{itemize}
    \item we consider regular blind periods of either 3.5 or 7 hours, occurring every 1 or 2 weeks, respectively, as representative of \textit{scheduled gaps};
    \item we assume a fixed duration of 3 days per \textit{unscheduled gap}, and model the interval $\Delta T$ between consecutive gaps as an exponential distribution
\begin{equation}
    \label{eq:gaps_dist}
    p(\Delta T \mid \lambda) = \lambda \, \exp\left[-\lambda \Delta T\right],
\end{equation}
where the rate parameter $\lambda$ is chosen to match an expected mission duty cycle of approximately $70\%$.
\end{itemize}

With this setup, we first analyze data corresponding to a total observation time of $T_{\rm obs} = 1$ year, divided into 2 weeks-long segments, and accounting only for scheduled gaps occurring every two weeks, each lasting for $T_{\rm gap} = 7 {\rm hr}$.  

\begin{figure*}
    \centering
    \includegraphics[width=2\columnwidth]{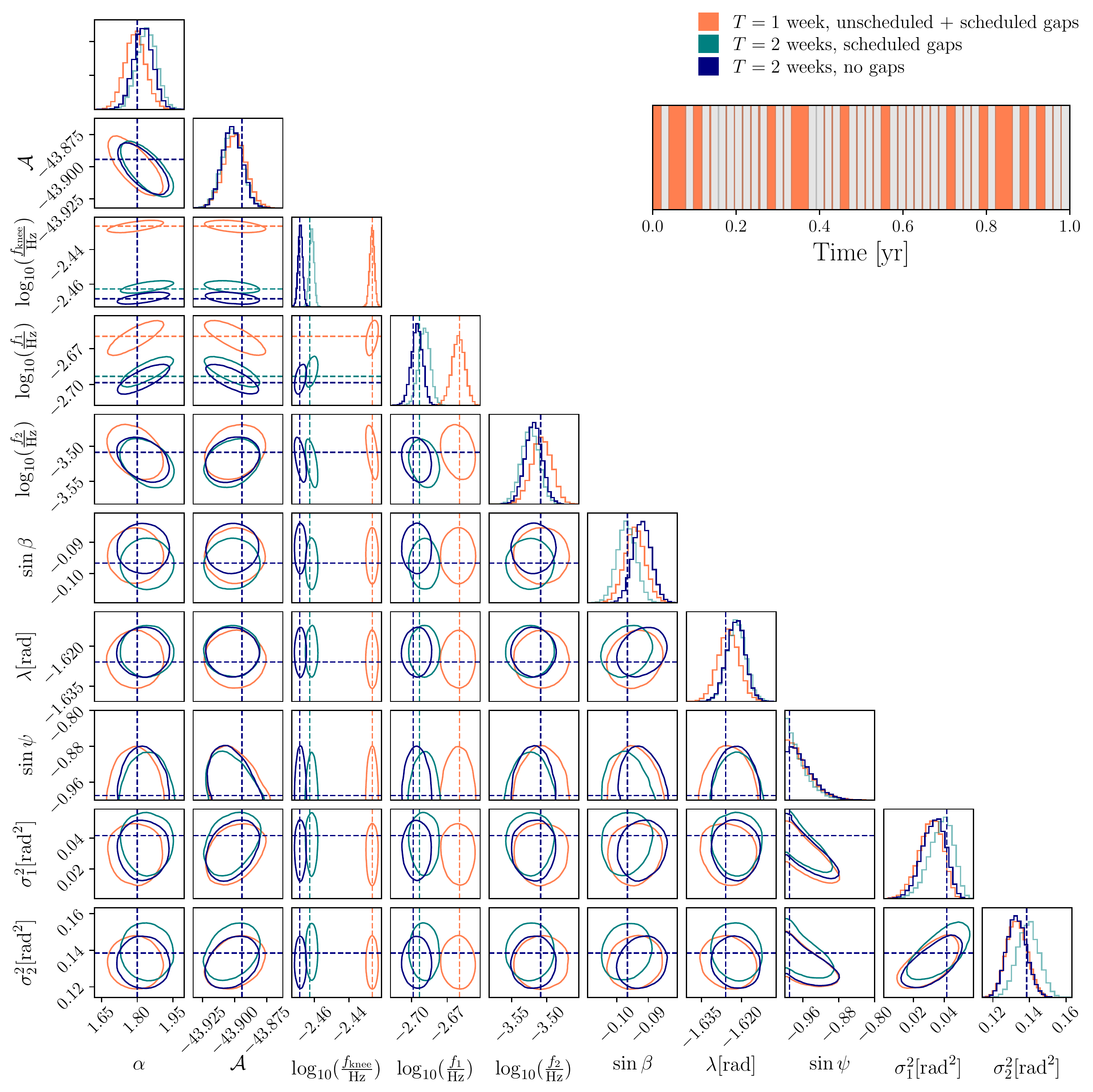}
    \caption{Posterior distribution of the Galactic foreground under different data–gap configurations.
    Blue contours correspond to the reference case with full data and no gaps. Teal contours include scheduled gaps of 7 hours every two weeks, while coral contours represent the most pessimistic case, with both scheduled and unscheduled gaps.  
    The top right inset illustrates the gaps arrangement, shown as coral vertical bars over a full 1 year timeline. In this scenario, the overall duty cycle is about $70\%$. 
    Dashed lines in the main panel denote the true injected values. 
    Note that the injected $\log_{10} f_{\rm{knee}}$ and $\log_{10} f_1$ are not equal across the three cases considered, as their value depend on the effective observation time $T_{\rm obs}$, which is reduced in the presence of gaps.}
    \label{fig:corner_chunk2}
\end{figure*}

In a second, more realistic scenario, we include both scheduled and unscheduled gaps. 
The former occur weekly with $T_{\rm gap} = 3.5 {\rm hr}$, while realisations of unscheduled gaps are generated according to the distribution in~\Cref{eq:gaps_dist}. 
As a consequence, the data segments have variable durations due to the irregular occurrence of unscheduled gaps. We adopt a conservative approach and retain only chunks with an effective duration of $T_{\rm chunk} = 1$ week, discarding all other data.
An example of such data segmentation is shown in the subplot of \Cref{fig:corner_chunk2}.

Therein, (i) we illustrate the chosen unscheduled gap realisation to achieve a target duty cycle close to $70\%$ and (ii) compare posterior distributions obtained in absence of gaps, with scheduled gaps only, and with both scheduled and unscheduled gaps.
For both scenarios, the presence of gaps does not compromise the reconstruction of either the Galactic foreground or the instrumental noise. 
We highlight that $f_{\rm{knee}}$ and $f_{1}$ take different injected values compared to the no-gap case, since the presence of gaps reduces the effective observation time, leading to modified values in~\Cref{eq:f1,eq:fknee}.
We highlight that for the instrumental noise model, we make the simplifying assumption of same amplitude level before and after each gap. 
This is, of course, a simplification, as the LISA noise level could vary at the onset of data taking after a gap.

\subsection{Extragalactic foreground \label{subsec:ef}}

We now study how a putative additional EF in the datastreams influences the time–frequency reconstruction of the GF. 
As discussed in~\Cref{subsec:lik_time_freq}, we inject it as an additional stationary, isotropic, Gaussian noise component. The corresponding PSD model is given by \cref{eq:extra_foreground}.
We perform inference using the sequential approach and two-weeks long segments. 
Specifically, we consider three alternative hypotheses:
\begin{itemize}[leftmargin=55pt,parsep=2pt,partopsep=6pt]
    \item[$\rm{GF}_{\rm{QS}} \cup \rm{EF}$]: we model consistently the stationary EF the quasi-stationary GF;
    \item[$\rm{GF}_{\rm{S}} \cup \rm{EF}$]: we model both EF and GF as stationary;
    \item[$\rm{GF}_{\rm{QS}}$]: we neglect the EF and model the GF as quasi-stationary.
\end{itemize}

For brevity, we omit from all three model names the instrumental noise, which is nonetheless jointly inferred upon. 
As discussed in \Cref{subsec:lik_time_freq}, the cutoff frequency of the EF spectrum occurs at approximately $40.2\mathrm{mHz}$, whereas the Galactic knee frequency, following the parametrization in \cref{eq:fknee}, spans from about $10\mathrm{mHz}$ down to $3.4\mathrm{mHz}$ over the course of a year. This clear spectral separation keeps the two components distinguishable. For this reason, we do not further infer on $f_{\rm cut}$.

The time evolution of the reconstruction for the $\rm{GF}_{\rm{QS}} \cup \rm{EF}$ model is illustrated by the ridgeplot in~\Cref{fig:ridgeline2}:
the posterior on the EF becomes informative starting from packet 6, indicating that after roughly 3 months we may be able to detect it with large confidence. 
We leave a more detailed study of the EF detectability as a function of the underlying assumptions to future work.

\begin{figure}
    \centering
    \includegraphics[width=1\columnwidth]{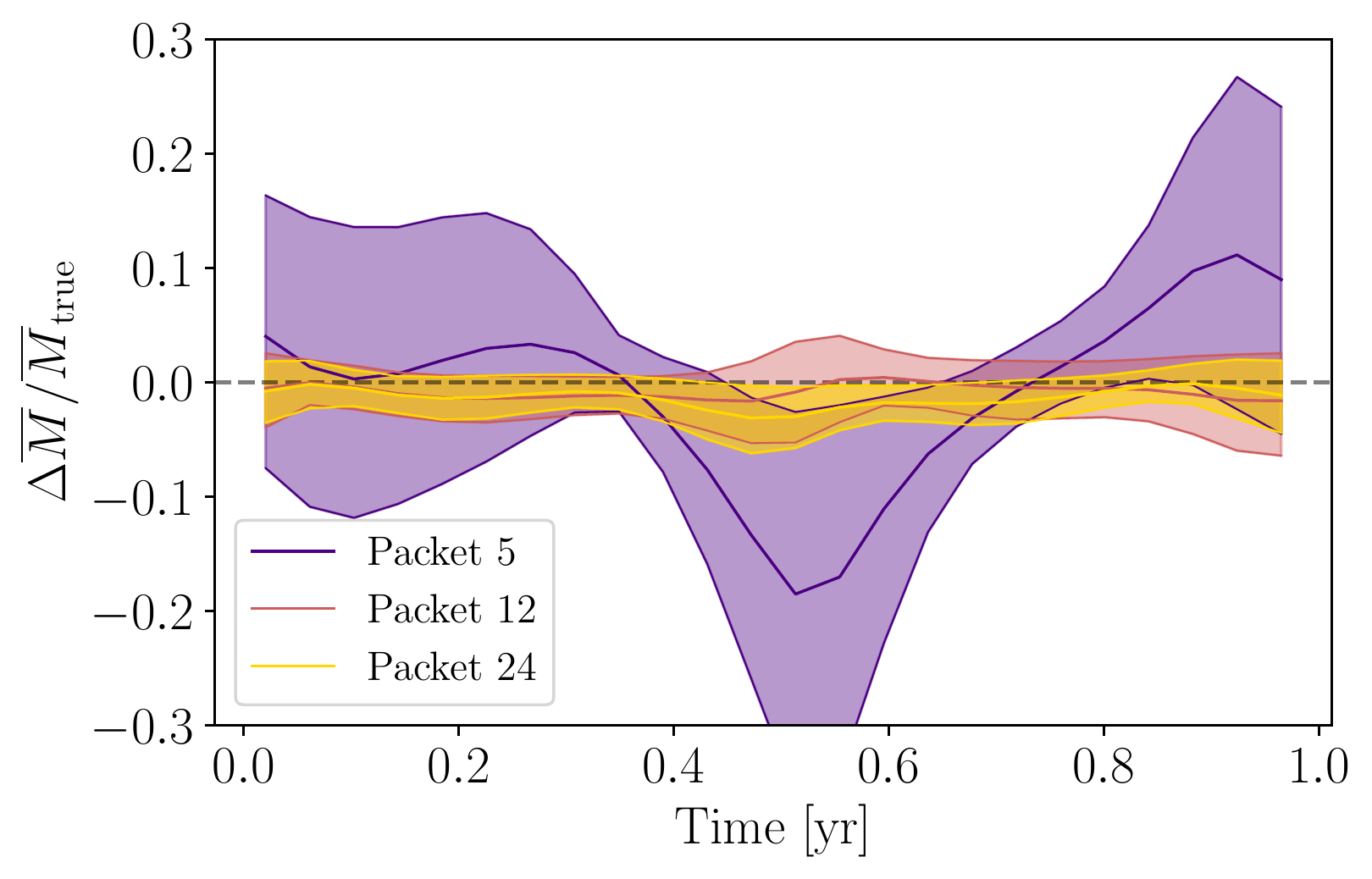}
    \includegraphics[width=1\columnwidth]{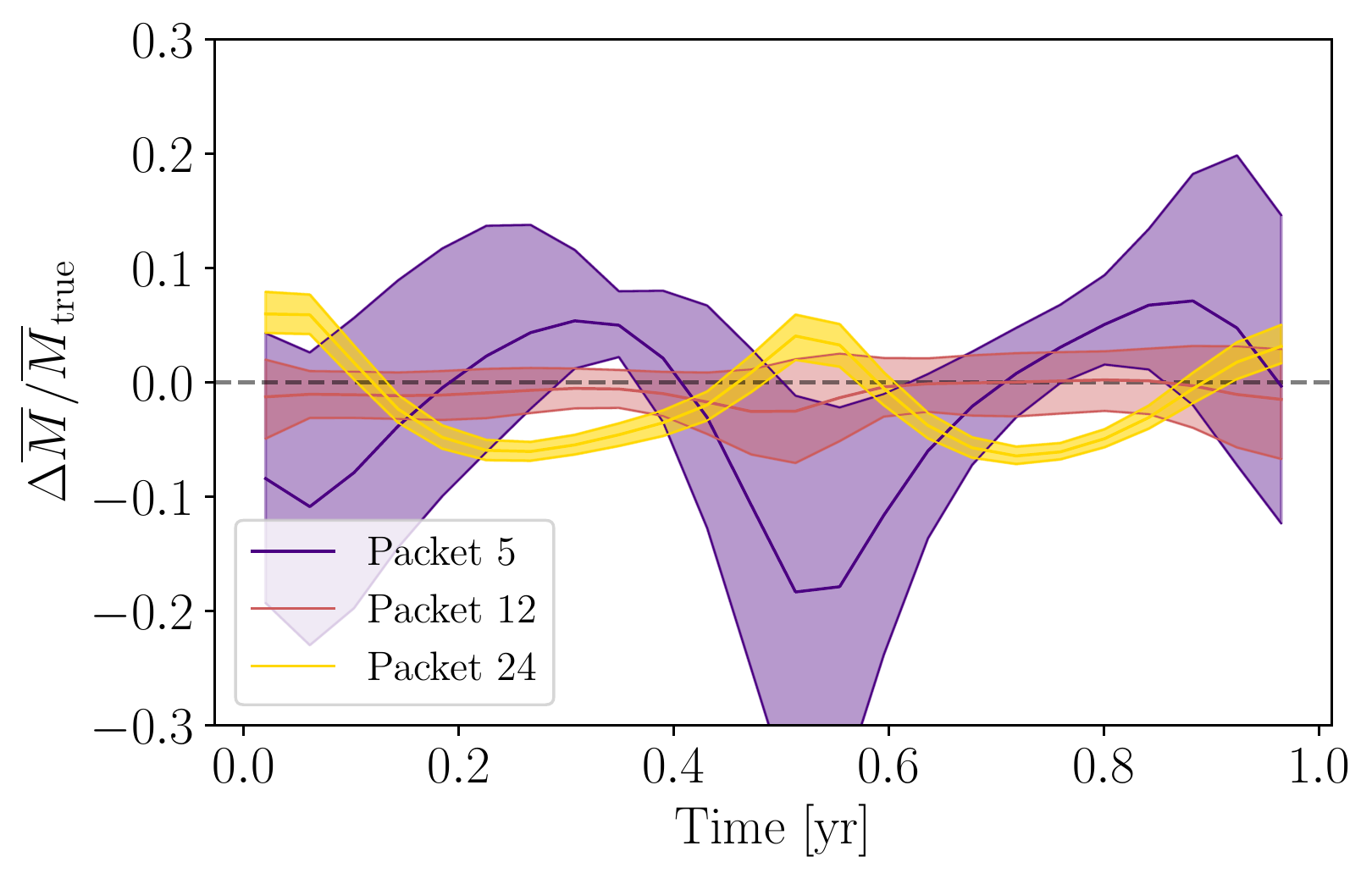}
    \caption{
    Quasi-stationary GF analysis on data containing an additional extra-Galactic foreground.
    (\textit{Top panel}) Posterior on the squared modulation amplitude $\overline{M}$, as introduced in~\Cref{eq:mbar}, with a model inferring on the EF.
    We plot the median and 90\% credible interval on the relative fractional error 
    $\Delta\overline{M}/\overline{M}_{\rm true}$ between the injected $\overline{M}_{\rm true}$ and reconstructed spectrum modulation.
    In purple (coral, yellow) we show results from the sequential analysis up to packets 5 (12,24), projected over the whole year.
    (\textit{Bottom panel}) Same as top panel, although with EF not accounted for in the model hypothesis.
    }
\label{fig:res_mod}
\end{figure}

We instead focus on the EF impact on the GF reconstruction: we observe that the presence of an EF does not compromise the reconstruction of the GF, provided the former is suitably incorporated into the data model. 
To demonstrate this we focus on the sequential analyses of the first 5, 12 and 24 packets, i.e. roughly corresponding to 2.5 months, 6 months, and 1 year of data.
We construct the posterior on the modulation envelope  and compute the fractional error relative to the injected value $M_{\rm true}$.
We therefore  plot the posterior on $\Delta M/M_{\rm true}$ over the whole mission first year --- as obtained assuming assuming the ${\rm GF}_{\rm QS} \cup {\rm EF}$ model --- in the top panel of \Cref{fig:res_mod}. 
We note that, even if constraints become available only once the corresponding data packets are acquired, their predictive effect is non-local in time, i.e. propagates back into past observations when analyzing deterministic sources in a global fit Gibbs-like scheme.

When only the first five packets (purple shaded area) are considered, the posterior predictive distribution over $\Delta M/M_{\rm true}$ is poorly constrained and marginally biased.
However, after roughly six months, an unbiased GF reconstruction emerges with nearly the same precision achieved after a full year of observation, namely at percent level.
Achieving such high precision in the measurement of the modulation is crucial for future prospects of constraining key parameters describing the Galactic geometry, such as the bulge and disk sizes.

\begin{figure}
    \centering
    \includegraphics[width=1\columnwidth]{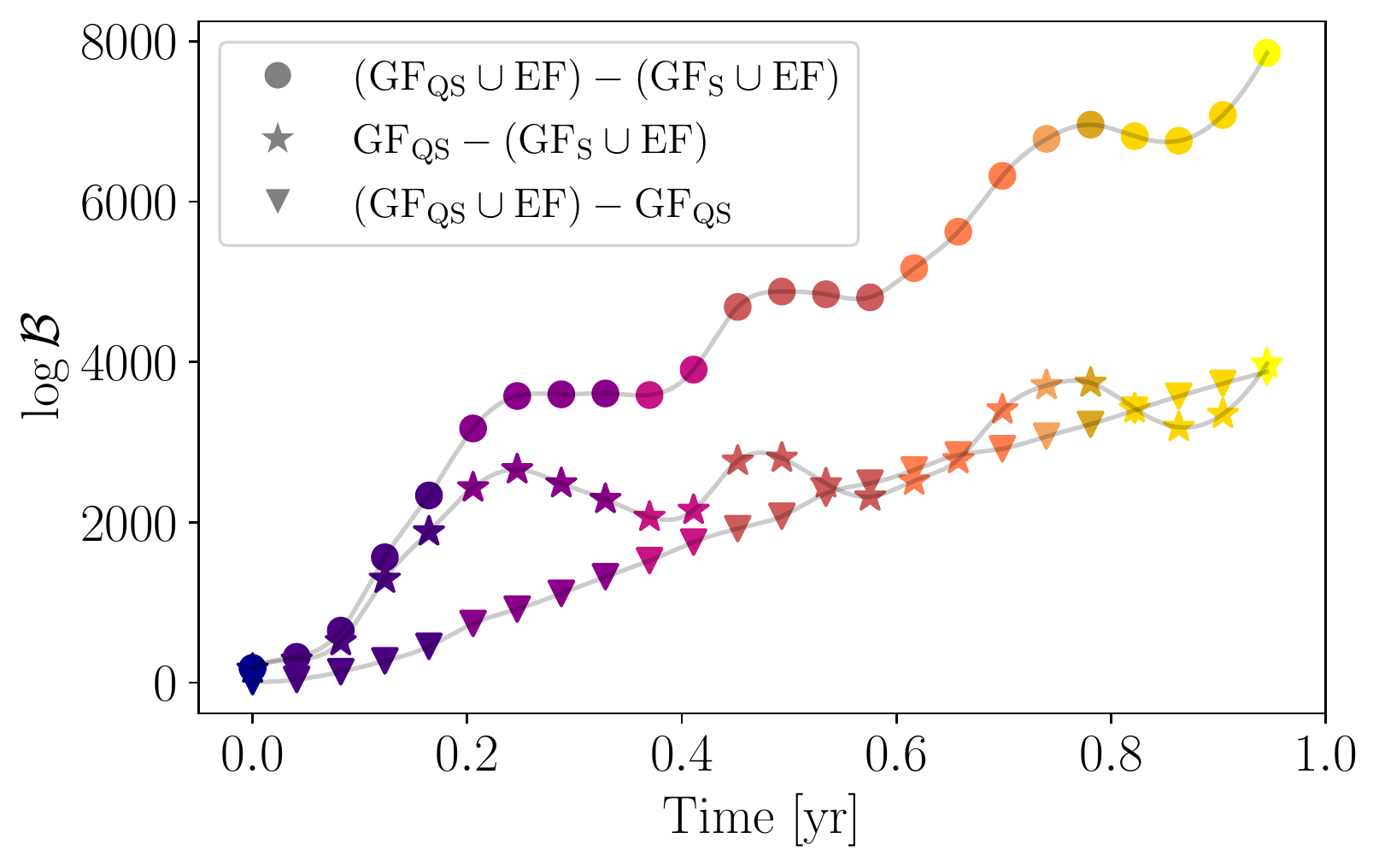}
    \caption{Evolution of the $\log$ Bayes factor between three considered hypotheses. The injected data include instrumental noise, GF, and EF. Circle (triangle) markers denote the log evidence-ratio of the full model against one not accounting for the GF quasistationarity (the presence of an EF).
    All three hypotheses account for instrumental noise.
    The information loss due to neglecting the GF quasi-stationarity is systematically larger than that induced by ignoring the EF. 
    The latter exhibits no oscillating behaviour as the unmodelled component is fully stationary.
    As shown by the starred markers, the difference oscillates in $T_{\rm obs}$ and has a shallower upward trend as compared to the full model, due to the stationary mismodelling of the GF and the unaccounted EF.
     }
\label{fig:bf_ef}
\end{figure}

\begin{figure*}
    \centering
    \includegraphics[width=2\columnwidth]{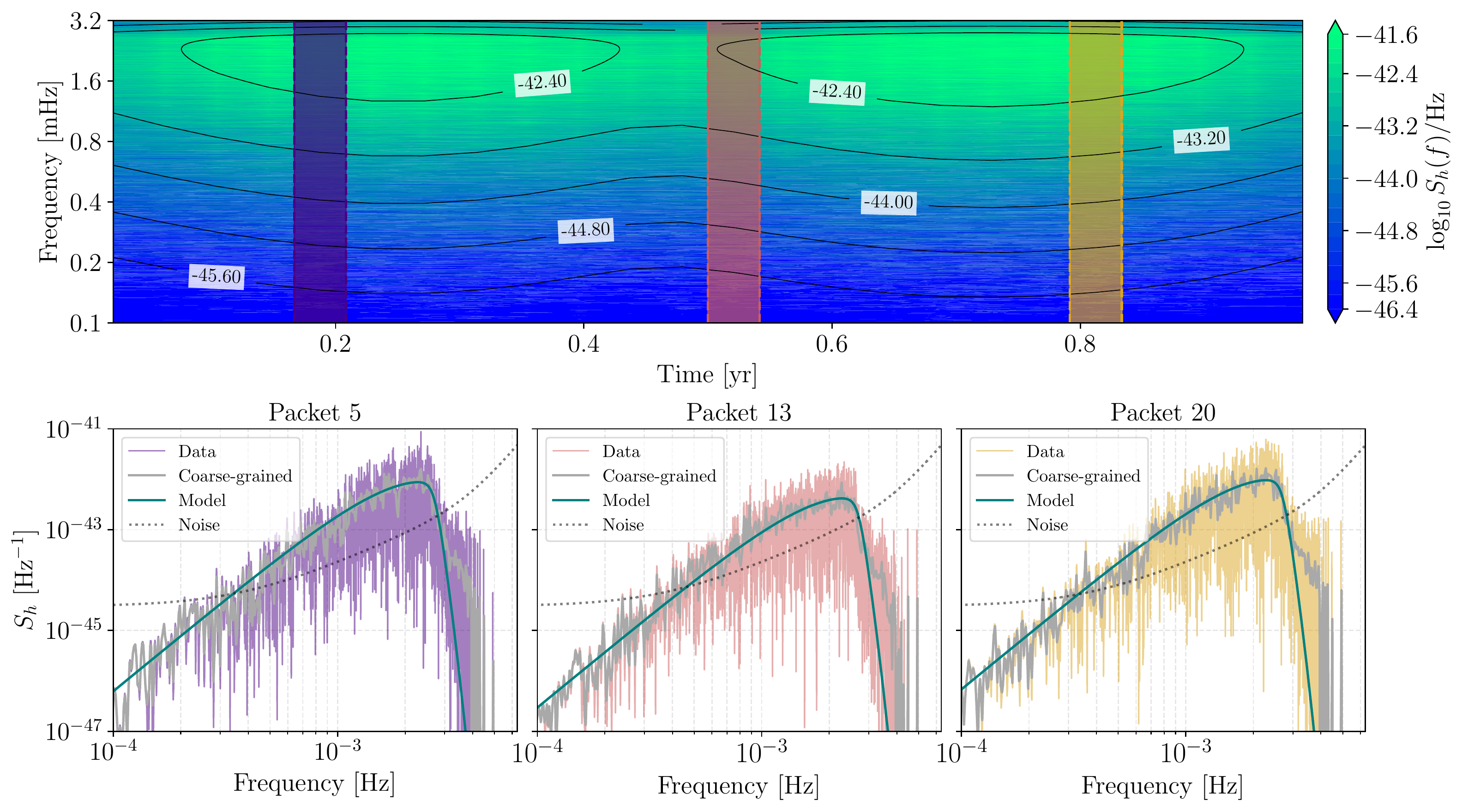}
    \caption{Spectrogram of the first year of  GF data from the \textit{Yorsh} data challenge. 
    The black annotated contours denote the median power spectral reconstruction from the posterior distribution of the quasi-stationary model. Colored rectangles highlight three representative data segments, whose PSDs are shown in the bottom panels. 
    Therein, each solid teal line represents the median PSD model posterior over the given segment, while the purple (coral, yellow) line denotes the PSD estimate obtained directly from data segment number 5 (13, 20). 
    Gray lines correspond to the coarse-grained data PSD, as described in~\Cref{sec:data_model}.
    For reference, dotted black lines denote the instrumental noise PSD level.
    }
\label{fig:yorsh}
\end{figure*}

Analogously, the bottom panel of \Cref{fig:res_mod} shows the same quantity obtained from inferences where the EF is neglected. As in the previous case, the modulation is initially poorly constrained. As more segments are collected, the reconstruction becomes nearly unbiased. However, as GF and EF spectra builds up signal-to-noise ratio over time, their cross-contamination becomes comparable to their respective posterior uncertainties. 
Therefore, after one year of observation, the modulation is severely biased if the EF is not accounted for. 
Our findings illustrate how cyclostationarity provides strong degeneracy breaking power, allowing the modulation—and thus potentially the structure of the Milky Way inferred from unresolved sources—to be characterized even after the first few month of observations. Nevertheless, considering predictions over a full year of data ---thereby leveraging the complete information encoded in the LISA response to anisotropic background--- is safer to achieve unbiased results, and fully breaking parameter degeneracies.

We conclude illustrating in~\Cref{fig:bf_ef} the Bayes factors between the three hypotheses introduced above.
Circled and triangular markers denote the log-evidence ratio when the GF cyclostationarity and the EF are ignored against the full model, respectively.
Notably, the log-evidence loss due to neglecting the GF quasi-stationarity is systematically larger than that induced by ignoring the EF. 
From a statistical perspective, it is therefore preferrable to neglect the EF rather than sacrifice the GF quasi-stationarity. 
As shown by the starred markers, the difference is oscillating upwards in $T_{\rm obs}$, a result of both mismodelling the GF quasistationarity and not accounting for the presence of a EF.

\subsection{Application to \textit{Yorsh} \label{subsec:yorsh}}
As a final validation step, we apply our framework to realistic observational data. Specifically, we analyze the Yorsh 1b dataset \cite{LDC1b}, which contains two years of time-domain data modelled as second-generation TDI variables $(X, Y, Z)$, downsampled to 5 second cadence. 
While this dataset was originally designed for testing the reconstruction of stellar-origin black hole binaries, its lego-like structure allows us to isolate data from instrumental noise and GF, and analyse them. 

We show in~\Cref{fig:yorsh} the GF reconstruction:
in the top panel the GF time-frequency spectrum obtained from the data is overlaid with contours of constant PSD, as posterior median level inferred by our model sequentially over each segment.
This visualization demonstrates our framework's ability to recover the characteristic time-frequency structure of the galactic signal. 
In addition, we focus on the spectral reconstruction of posterior predictive PSDs from the quasi-stationary model for packets 5, 13, and 20, corresponding to an accumulation of 10, 26 and 40 weeks of observation, respectively. 
Although residual spurious power at frequencies above 3 mHz is not captured --- likely because of near-threshold unresolved DWDs--- the PSD reconstruction at lower frequencies is robust and unbiased, despite the increased complexity of the data with respect to our model.

\section{Conclusion}
\label{sec:conclusion}
In this manuscript, we presented an extensive analysis of the Galactic foreground generated by unresolved white dwarf binaries in LISA, focusing on the first year of data. 
This astrophysical noise component represents one of the major challenges for LISA data analysis, and it heavily influences the global fit likelihood model.
Disentangling it from instrumental noise, is therefore paramount.

Building on the \texttt{bahamas} code introduced in~\cite{2025arXiv250622542P}, we addressed a number of challenges, mainly focusing on a global-fit deployment of our methodology. 
By exploiting the time–frequency model for the data, we obtained robust results across a range of configurations. 
In particular, we found that using averaged data combined with the NUTS sampler significantly reduces the computational cost of the analysis, yielding a speed-up factor of about $30$ with respect to standard approaches based on the Whittle likelihood and NS sampling.

We then quantified the reconstruction of Galactic foreground parameters, investigating robustness against additional features, i.e. data gaps. 
First, we investigated the impact of data gaps considering regularly scheduled and unscheduled interruptions. 
Even in the most extreme case, with almost $\sim30\%$ loss on the observing time, the reconstruction of the GF remains largely unaffected.
Second, we studied the effect of a moderately bright DWD background of extragalactic origin. 
If not accounted for in the model, and despite being isotropic and stationary, neglecting it induces systematic biases in the GF spectrum, in particular on the Galaxy spatial distribution parameters. 
Our analysis shows that it is preferable to neglect this extra component rather than include it while treating the GF as stationary, underscoring the importance of adopting a quasi-stationary signal model.

While our results demonstrate the flexibility and modularity of \texttt{bahamas}, several extensions are envisioned: the explicit inclusion of instrumental noise non-stationarity, which may arise from the LISA constellation breathing or from intrinsic time-dependent variations in the instrumental noise level; its full integration in a global-fit framework, thus accounting for deterministic resolvable signals without the iterative subtraction approximation; the further extension of the data model to higher correlation functions beyond the power spectrum, to capture non-Gaussian features of the Galactic foreground.
These aspects will be addressed in future investigations.




\vspace{-1.0em}
\begin{acknowledgments}
The authors are grateful to J.~Gair, C.~Moore, R.~Meyer, M.~Pieroni, Q.Baghi, A. Criswell, N. Karnesis, L.Valbusa Dall'Armi, G. Astorino, A.~Spadaro, C.~Caprini, G.~Nardini, and all members of the LISA Distributed Data Processing Center for valuable inputs.
R.B. acknowledges financial support by the Italian Space Agency grant Phase B2/C activity for LISA mission, Agreement n.2024-NAZ-0102/PE.
A.S. acknowledges financial support provided under the European Union’s H2020 ERC Agreement: 818691) and Advanced
Grant “PINGU” (Grant Agreement: 101142079).
Computational resources were provided by University of Birmingham BlueBEAR High Performance Computing facility and CINECA through EuroHPC Benchmark access call grant EHPC-BEN-2025B08-042.
\paragraph*{\textbf{Data and code availability}} The data supporting findings in this manuscript are openly available through the Zenodo record~\cite{bahamasgf-release} and at the 
\href{https://github.com/RiccardoBuscicchio/bahamas-galactic-foreground-release}{GitHub release}.
The \texttt{bahamas} code version used in this work is tagged and available as the \href{https://github.com/FedericoPozzoli/bahamas/releases/tag/v1.0.0-gf}{\texttt{v1.0.0-gf}} release on GitHub.
\vfill
\end{acknowledgments}

\bibliographystyle{apsrev4-2}
\bibliography{main}

\begin{thebibliography}{42}%
\makeatletter
\providecommand \@ifxundefined [1]{%
 \@ifx{#1\undefined}
}%
\providecommand \@ifnum [1]{%
 \ifnum #1\expandafter \@firstoftwo
 \else \expandafter \@secondoftwo
 \fi
}%
\providecommand \@ifx [1]{%
 \ifx #1\expandafter \@firstoftwo
 \else \expandafter \@secondoftwo
 \fi
}%
\providecommand \natexlab [1]{#1}%
\providecommand \enquote  [1]{``#1''}%
\providecommand \bibnamefont  [1]{#1}%
\providecommand \bibfnamefont [1]{#1}%
\providecommand \citenamefont [1]{#1}%
\providecommand \href@noop [0]{\@secondoftwo}%
\providecommand \href [0]{\begingroup \@sanitize@url \@href}%
\providecommand \@href[1]{\@@startlink{#1}\@@href}%
\providecommand \@@href[1]{\endgroup#1\@@endlink}%
\providecommand \@sanitize@url [0]{\catcode `\\12\catcode `\$12\catcode
  `\&12\catcode `\#12\catcode `\^12\catcode `\_12\catcode `\%12\relax}%
\providecommand \@@startlink[1]{}%
\providecommand \@@endlink[0]{}%
\providecommand \url  [0]{\begingroup\@sanitize@url \@url }%
\providecommand \@url [1]{\endgroup\@href {#1}{\urlprefix }}%
\providecommand \urlprefix  [0]{URL }%
\providecommand \Eprint [0]{\href }%
\providecommand \doibase [0]{https://doi.org/}%
\providecommand \selectlanguage [0]{\@gobble}%
\providecommand \bibinfo  [0]{\@secondoftwo}%
\providecommand \bibfield  [0]{\@secondoftwo}%
\providecommand \translation [1]{[#1]}%
\providecommand \BibitemOpen [0]{}%
\providecommand \bibitemStop [0]{}%
\providecommand \bibitemNoStop [0]{.\EOS\space}%
\providecommand \EOS [0]{\spacefactor3000\relax}%
\providecommand \BibitemShut  [1]{\csname bibitem#1\endcsname}%
\let\auto@bib@innerbib\@empty
\bibitem [{\citenamefont {Amaro-Seoane}\ \emph {et~al.}(2017)\citenamefont
  {Amaro-Seoane} \emph {et~al.}}]{2017arXiv170200786A}%
  \BibitemOpen
  \bibfield  {author} {\bibinfo {author} {\bibfnamefont {P.}~\bibnamefont
  {Amaro-Seoane}} \emph {et~al.},\ }\href
  {https://doi.org/10.48550/arXiv.1702.00786} {\bibfield  {journal} {\bibinfo
  {journal} {arXiv e-prints}\ ,\ \bibinfo {eid} {arXiv:1702.00786}} (\bibinfo
  {year} {2017})},\ \Eprint {https://arxiv.org/abs/1702.00786}
  {arXiv:1702.00786 [astro-ph.IM]} \BibitemShut {NoStop}%
\bibitem [{\citenamefont {Colpi}\ and\ \citenamefont
  {others.}(2024)}]{Colpi:2024}%
  \BibitemOpen
  \bibfield  {author} {\bibinfo {author} {\bibfnamefont {M.}~\bibnamefont
  {Colpi}}\ and\ \bibinfo {author} {\bibnamefont {others.}},\ }\href
  {https://doi.org/10.48550/arXiv.2402.07571} {\bibfield  {journal} {\bibinfo
  {journal} {arXiv e-prints}\ ,\ \bibinfo {eid} {arXiv:2402.07571}} (\bibinfo
  {year} {2024})},\ \Eprint {https://arxiv.org/abs/2402.07571}
  {arXiv:2402.07571 [astro-ph.CO]} \BibitemShut {NoStop}%
\bibitem [{\citenamefont {{Littenberg}}\ and\ \citenamefont
  {{Cornish}}(2023)}]{2023PhRvD.107f3004L}%
  \BibitemOpen
  \bibfield  {author} {\bibinfo {author} {\bibfnamefont {T.~B.}\ \bibnamefont
  {{Littenberg}}}\ and\ \bibinfo {author} {\bibfnamefont {N.~J.}\ \bibnamefont
  {{Cornish}}},\ }\href {https://doi.org/10.1103/PhysRevD.107.063004}
  {\bibfield  {journal} {\bibinfo  {journal} {\prd}\ }\textbf {\bibinfo
  {volume} {107}},\ \bibinfo {eid} {063004} (\bibinfo {year} {2023})},\ \Eprint
  {https://arxiv.org/abs/2301.03673} {arXiv:2301.03673 [gr-qc]} \BibitemShut
  {NoStop}%
\bibitem [{\citenamefont {{Katz}}\ \emph {et~al.}(2025)\citenamefont {{Katz}},
  \citenamefont {{Karnesis}}, \citenamefont {{Korsakova}}, \citenamefont
  {{Gair}},\ and\ \citenamefont {{Stergioulas}}}]{2025PhRvD.111b4060K}%
  \BibitemOpen
  \bibfield  {author} {\bibinfo {author} {\bibfnamefont {M.~L.}\ \bibnamefont
  {{Katz}}}, \bibinfo {author} {\bibfnamefont {N.}~\bibnamefont {{Karnesis}}},
  \bibinfo {author} {\bibfnamefont {N.}~\bibnamefont {{Korsakova}}}, \bibinfo
  {author} {\bibfnamefont {J.~R.}\ \bibnamefont {{Gair}}},\ and\ \bibinfo
  {author} {\bibfnamefont {N.}~\bibnamefont {{Stergioulas}}},\ }\href
  {https://doi.org/10.1103/PhysRevD.111.024060} {\bibfield  {journal} {\bibinfo
   {journal} {\prd}\ }\textbf {\bibinfo {volume} {111}},\ \bibinfo {eid}
  {024060} (\bibinfo {year} {2025})},\ \Eprint
  {https://arxiv.org/abs/2405.04690} {arXiv:2405.04690 [gr-qc]} \BibitemShut
  {NoStop}%
\bibitem [{\citenamefont {{Deng}}\ \emph {et~al.}(2025)\citenamefont {{Deng}},
  \citenamefont {{Babak}}, \citenamefont {{Le Jeune}}, \citenamefont
  {{Marsat}}, \citenamefont {{Plagnol}},\ and\ \citenamefont
  {{Sartirana}}}]{2025PhRvD.111j3014D}%
  \BibitemOpen
  \bibfield  {author} {\bibinfo {author} {\bibfnamefont {S.}~\bibnamefont
  {{Deng}}}, \bibinfo {author} {\bibfnamefont {S.}~\bibnamefont {{Babak}}},
  \bibinfo {author} {\bibfnamefont {M.}~\bibnamefont {{Le Jeune}}}, \bibinfo
  {author} {\bibfnamefont {S.}~\bibnamefont {{Marsat}}}, \bibinfo {author}
  {\bibfnamefont {{\'E}.}~\bibnamefont {{Plagnol}}},\ and\ \bibinfo {author}
  {\bibfnamefont {A.}~\bibnamefont {{Sartirana}}},\ }\href
  {https://doi.org/10.1103/PhysRevD.111.103014} {\bibfield  {journal} {\bibinfo
   {journal} {\prd}\ }\textbf {\bibinfo {volume} {111}},\ \bibinfo {eid}
  {103014} (\bibinfo {year} {2025})},\ \Eprint
  {https://arxiv.org/abs/2501.10277} {arXiv:2501.10277 [gr-qc]} \BibitemShut
  {NoStop}%
\bibitem [{\citenamefont {{Strub}}\ \emph {et~al.}(2024)\citenamefont
  {{Strub}}, \citenamefont {{Ferraioli}}, \citenamefont {{Schmelzbach}},
  \citenamefont {{St{\"a}hler}},\ and\ \citenamefont
  {{Giardini}}}]{2024PhRvD.110b4005S}%
  \BibitemOpen
  \bibfield  {author} {\bibinfo {author} {\bibfnamefont {S.~H.}\ \bibnamefont
  {{Strub}}}, \bibinfo {author} {\bibfnamefont {L.}~\bibnamefont
  {{Ferraioli}}}, \bibinfo {author} {\bibfnamefont {C.}~\bibnamefont
  {{Schmelzbach}}}, \bibinfo {author} {\bibfnamefont {S.~C.}\ \bibnamefont
  {{St{\"a}hler}}},\ and\ \bibinfo {author} {\bibfnamefont {D.}~\bibnamefont
  {{Giardini}}},\ }\href {https://doi.org/10.1103/PhysRevD.110.024005}
  {\bibfield  {journal} {\bibinfo  {journal} {\prd}\ }\textbf {\bibinfo
  {volume} {110}},\ \bibinfo {eid} {024005} (\bibinfo {year} {2024})},\ \Eprint
  {https://arxiv.org/abs/2403.15318} {arXiv:2403.15318 [gr-qc]} \BibitemShut
  {NoStop}%
\bibitem [{\citenamefont {{Nelemans}}(2009)}]{2009CQGra..26i4030N}%
  \BibitemOpen
  \bibfield  {author} {\bibinfo {author} {\bibfnamefont {G.}~\bibnamefont
  {{Nelemans}}},\ }\href {https://doi.org/10.1088/0264-9381/26/9/094030}
  {\bibfield  {journal} {\bibinfo  {journal} {Classical and Quantum Gravity}\
  }\textbf {\bibinfo {volume} {26}},\ \bibinfo {eid} {094030} (\bibinfo {year}
  {2009})},\ \Eprint {https://arxiv.org/abs/0901.1778} {arXiv:0901.1778
  [astro-ph.SR]} \BibitemShut {NoStop}%
\bibitem [{\citenamefont {{Perego}}\ \emph {et~al.}(2025)\citenamefont
  {{Perego}}, \citenamefont {{Bonetti}}, \citenamefont {{Sesana}},
  \citenamefont {{Toonen}},\ and\ \citenamefont
  {{Korol}}}]{2025arXiv251018695P}%
  \BibitemOpen
  \bibfield  {author} {\bibinfo {author} {\bibfnamefont {A.}~\bibnamefont
  {{Perego}}}, \bibinfo {author} {\bibfnamefont {M.}~\bibnamefont {{Bonetti}}},
  \bibinfo {author} {\bibfnamefont {A.}~\bibnamefont {{Sesana}}}, \bibinfo
  {author} {\bibfnamefont {S.}~\bibnamefont {{Toonen}}},\ and\ \bibinfo
  {author} {\bibfnamefont {V.}~\bibnamefont {{Korol}}},\ }\href
  {https://doi.org/10.48550/arXiv.2510.18695} {\bibfield  {journal} {\bibinfo
  {journal} {arXiv e-prints}\ ,\ \bibinfo {eid} {arXiv:2510.18695}} (\bibinfo
  {year} {2025})},\ \Eprint {https://arxiv.org/abs/2510.18695}
  {arXiv:2510.18695 [astro-ph.HE]} \BibitemShut {NoStop}%
\bibitem [{\citenamefont {{Boileau}}\ \emph {et~al.}(2025)\citenamefont
  {{Boileau}}, \citenamefont {{Bruel}}, \citenamefont {{Toubiana}},
  \citenamefont {{Lamberts}},\ and\ \citenamefont
  {{Christensen}}}]{2025arXiv250618390B}%
  \BibitemOpen
  \bibfield  {author} {\bibinfo {author} {\bibfnamefont {G.}~\bibnamefont
  {{Boileau}}}, \bibinfo {author} {\bibfnamefont {T.}~\bibnamefont {{Bruel}}},
  \bibinfo {author} {\bibfnamefont {A.}~\bibnamefont {{Toubiana}}}, \bibinfo
  {author} {\bibfnamefont {A.}~\bibnamefont {{Lamberts}}},\ and\ \bibinfo
  {author} {\bibfnamefont {N.}~\bibnamefont {{Christensen}}},\ }\href
  {https://doi.org/10.48550/arXiv.2506.18390} {\bibfield  {journal} {\bibinfo
  {journal} {arXiv e-prints}\ ,\ \bibinfo {eid} {arXiv:2506.18390}} (\bibinfo
  {year} {2025})},\ \Eprint {https://arxiv.org/abs/2506.18390}
  {arXiv:2506.18390 [gr-qc]} \BibitemShut {NoStop}%
\bibitem [{\citenamefont {{Hofman}}\ and\ \citenamefont
  {{Nelemans}}(2024)}]{2024A&A...691A.261H}%
  \BibitemOpen
  \bibfield  {author} {\bibinfo {author} {\bibfnamefont {S.}~\bibnamefont
  {{Hofman}}}\ and\ \bibinfo {author} {\bibfnamefont {G.}~\bibnamefont
  {{Nelemans}}},\ }\href {https://doi.org/10.1051/0004-6361/202451510}
  {\bibfield  {journal} {\bibinfo  {journal} {\aap}\ }\textbf {\bibinfo
  {volume} {691}},\ \bibinfo {eid} {A261} (\bibinfo {year} {2024})},\ \Eprint
  {https://arxiv.org/abs/2407.10642} {arXiv:2407.10642 [astro-ph.HE]}
  \BibitemShut {NoStop}%
\bibitem [{\citenamefont {{Staelens}}\ and\ \citenamefont
  {{Nelemans}}(2024)}]{2024A&A...683A.139S}%
  \BibitemOpen
  \bibfield  {author} {\bibinfo {author} {\bibfnamefont {S.}~\bibnamefont
  {{Staelens}}}\ and\ \bibinfo {author} {\bibfnamefont {G.}~\bibnamefont
  {{Nelemans}}},\ }\href {https://doi.org/10.1051/0004-6361/202348429}
  {\bibfield  {journal} {\bibinfo  {journal} {\aap}\ }\textbf {\bibinfo
  {volume} {683}},\ \bibinfo {eid} {A139} (\bibinfo {year} {2024})},\ \Eprint
  {https://arxiv.org/abs/2310.19448} {arXiv:2310.19448 [astro-ph.HE]}
  \BibitemShut {NoStop}%
\bibitem [{\citenamefont {{Buscicchio}}\ \emph {et~al.}(2025)\citenamefont
  {{Buscicchio}}, \citenamefont {{Klein}}, \citenamefont {{Korol}},
  \citenamefont {{Di Renzo}}, \citenamefont {{Moore}}, \citenamefont
  {{Gerosa}},\ and\ \citenamefont {{Carzaniga}}}]{2025EPJC...85..887B}%
  \BibitemOpen
  \bibfield  {author} {\bibinfo {author} {\bibfnamefont {R.}~\bibnamefont
  {{Buscicchio}}}, \bibinfo {author} {\bibfnamefont {A.}~\bibnamefont
  {{Klein}}}, \bibinfo {author} {\bibfnamefont {V.}~\bibnamefont {{Korol}}},
  \bibinfo {author} {\bibfnamefont {F.}~\bibnamefont {{Di Renzo}}}, \bibinfo
  {author} {\bibfnamefont {C.~J.}\ \bibnamefont {{Moore}}}, \bibinfo {author}
  {\bibfnamefont {D.}~\bibnamefont {{Gerosa}}},\ and\ \bibinfo {author}
  {\bibfnamefont {A.}~\bibnamefont {{Carzaniga}}},\ }\href
  {https://doi.org/10.1140/epjc/s10052-025-14616-w} {\bibfield  {journal}
  {\bibinfo  {journal} {European Physical Journal C}\ }\textbf {\bibinfo
  {volume} {85}},\ \bibinfo {eid} {887} (\bibinfo {year} {2025})},\ \Eprint
  {https://arxiv.org/abs/2410.08263} {arXiv:2410.08263 [astro-ph.HE]}
  \BibitemShut {NoStop}%
\bibitem [{\citenamefont {{Rosati}}\ and\ \citenamefont
  {{Littenberg}}(2024)}]{2024arXiv241017180R}%
  \BibitemOpen
  \bibfield  {author} {\bibinfo {author} {\bibfnamefont {R.}~\bibnamefont
  {{Rosati}}}\ and\ \bibinfo {author} {\bibfnamefont {T.~B.}\ \bibnamefont
  {{Littenberg}}},\ }\href {https://doi.org/10.48550/arXiv.2410.17180}
  {\bibfield  {journal} {\bibinfo  {journal} {arXiv e-prints}\ ,\ \bibinfo
  {eid} {arXiv:2410.17180}} (\bibinfo {year} {2024})},\ \Eprint
  {https://arxiv.org/abs/2410.17180} {arXiv:2410.17180 [gr-qc]} \BibitemShut
  {NoStop}%
\bibitem [{\citenamefont {{Karnesis}}\ \emph {et~al.}(2025)\citenamefont
  {{Karnesis}}, \citenamefont {{Sasli}}, \citenamefont {{Buscicchio}},\ and\
  \citenamefont {{Stergioulas}}}]{2025PhRvD.111b2005K}%
  \BibitemOpen
  \bibfield  {author} {\bibinfo {author} {\bibfnamefont {N.}~\bibnamefont
  {{Karnesis}}}, \bibinfo {author} {\bibfnamefont {A.}~\bibnamefont {{Sasli}}},
  \bibinfo {author} {\bibfnamefont {R.}~\bibnamefont {{Buscicchio}}},\ and\
  \bibinfo {author} {\bibfnamefont {N.}~\bibnamefont {{Stergioulas}}},\ }\href
  {https://doi.org/10.1103/PhysRevD.111.022005} {\bibfield  {journal} {\bibinfo
   {journal} {\prd}\ }\textbf {\bibinfo {volume} {111}},\ \bibinfo {eid}
  {022005} (\bibinfo {year} {2025})},\ \Eprint
  {https://arxiv.org/abs/2410.14354} {arXiv:2410.14354 [gr-qc]} \BibitemShut
  {NoStop}%
\bibitem [{\citenamefont {{Pozzoli}}\ \emph
  {et~al.}(2025{\natexlab{a}})\citenamefont {{Pozzoli}}, \citenamefont
  {{Buscicchio}}, \citenamefont {{Klein}}, \citenamefont {{Korol}},
  \citenamefont {{Sesana}},\ and\ \citenamefont
  {{Haardt}}}]{2025PhRvD.111f3005P}%
  \BibitemOpen
  \bibfield  {author} {\bibinfo {author} {\bibfnamefont {F.}~\bibnamefont
  {{Pozzoli}}}, \bibinfo {author} {\bibfnamefont {R.}~\bibnamefont
  {{Buscicchio}}}, \bibinfo {author} {\bibfnamefont {A.}~\bibnamefont
  {{Klein}}}, \bibinfo {author} {\bibfnamefont {V.}~\bibnamefont {{Korol}}},
  \bibinfo {author} {\bibfnamefont {A.}~\bibnamefont {{Sesana}}},\ and\
  \bibinfo {author} {\bibfnamefont {F.}~\bibnamefont {{Haardt}}},\ }\href
  {https://doi.org/10.1103/PhysRevD.111.063005} {\bibfield  {journal} {\bibinfo
   {journal} {\prd}\ }\textbf {\bibinfo {volume} {111}},\ \bibinfo {eid}
  {063005} (\bibinfo {year} {2025}{\natexlab{a}})},\ \Eprint
  {https://arxiv.org/abs/2410.08274} {arXiv:2410.08274 [astro-ph.GA]}
  \BibitemShut {NoStop}%
\bibitem [{\citenamefont {{Rieck}}\ \emph {et~al.}(2024)\citenamefont
  {{Rieck}}, \citenamefont {{Criswell}}, \citenamefont {{Korol}}, \citenamefont
  {{Keim}}, \citenamefont {{Bloom}},\ and\ \citenamefont
  {{Mandic}}}]{2024MNRAS.531.2642R}%
  \BibitemOpen
  \bibfield  {author} {\bibinfo {author} {\bibfnamefont {S.}~\bibnamefont
  {{Rieck}}}, \bibinfo {author} {\bibfnamefont {A.~W.}\ \bibnamefont
  {{Criswell}}}, \bibinfo {author} {\bibfnamefont {V.}~\bibnamefont {{Korol}}},
  \bibinfo {author} {\bibfnamefont {M.~A.}\ \bibnamefont {{Keim}}}, \bibinfo
  {author} {\bibfnamefont {M.}~\bibnamefont {{Bloom}}},\ and\ \bibinfo {author}
  {\bibfnamefont {V.}~\bibnamefont {{Mandic}}},\ }\href
  {https://doi.org/10.1093/mnras/stae1283} {\bibfield  {journal} {\bibinfo
  {journal} {\mnras}\ }\textbf {\bibinfo {volume} {531}},\ \bibinfo {pages}
  {2642} (\bibinfo {year} {2024})},\ \Eprint {https://arxiv.org/abs/2308.12437}
  {arXiv:2308.12437 [astro-ph.IM]} \BibitemShut {NoStop}%
\bibitem [{\citenamefont {{Digman}}\ and\ \citenamefont
  {{Cornish}}(2022)}]{2022ApJ...940...10D}%
  \BibitemOpen
  \bibfield  {author} {\bibinfo {author} {\bibfnamefont {M.~C.}\ \bibnamefont
  {{Digman}}}\ and\ \bibinfo {author} {\bibfnamefont {N.~J.}\ \bibnamefont
  {{Cornish}}},\ }\href {https://doi.org/10.3847/1538-4357/ac9139} {\bibfield
  {journal} {\bibinfo  {journal} {\apj}\ }\textbf {\bibinfo {volume} {940}},\
  \bibinfo {eid} {10} (\bibinfo {year} {2022})},\ \Eprint
  {https://arxiv.org/abs/2206.14813} {arXiv:2206.14813 [astro-ph.IM]}
  \BibitemShut {NoStop}%
\bibitem [{\citenamefont {{Criswell}}\ \emph
  {et~al.}(2025{\natexlab{a}})\citenamefont {{Criswell}}, \citenamefont
  {{Rieck}},\ and\ \citenamefont {{Mandic}}}]{2025PhRvD.111b3025C}%
  \BibitemOpen
  \bibfield  {author} {\bibinfo {author} {\bibfnamefont {A.~W.}\ \bibnamefont
  {{Criswell}}}, \bibinfo {author} {\bibfnamefont {S.}~\bibnamefont
  {{Rieck}}},\ and\ \bibinfo {author} {\bibfnamefont {V.}~\bibnamefont
  {{Mandic}}},\ }\href {https://doi.org/10.1103/PhysRevD.111.023025} {\bibfield
   {journal} {\bibinfo  {journal} {\prd}\ }\textbf {\bibinfo {volume} {111}},\
  \bibinfo {eid} {023025} (\bibinfo {year} {2025}{\natexlab{a}})},\ \Eprint
  {https://arxiv.org/abs/2410.23260} {arXiv:2410.23260 [astro-ph.IM]}
  \BibitemShut {NoStop}%
\bibitem [{\citenamefont {{Hindmarsh}}\ \emph {et~al.}(2025)\citenamefont
  {{Hindmarsh}}, \citenamefont {{Hooper}}, \citenamefont {{Minkkinen}},\ and\
  \citenamefont {{Weir}}}]{2025JCAP...04..052H}%
  \BibitemOpen
  \bibfield  {author} {\bibinfo {author} {\bibfnamefont {M.}~\bibnamefont
  {{Hindmarsh}}}, \bibinfo {author} {\bibfnamefont {D.~C.}\ \bibnamefont
  {{Hooper}}}, \bibinfo {author} {\bibfnamefont {T.}~\bibnamefont
  {{Minkkinen}}},\ and\ \bibinfo {author} {\bibfnamefont {D.~J.}\ \bibnamefont
  {{Weir}}},\ }\href {https://doi.org/10.1088/1475-7516/2025/04/052} {\bibfield
   {journal} {\bibinfo  {journal} {\jcap}\ }\textbf {\bibinfo {volume}
  {2025}},\ \bibinfo {eid} {052} (\bibinfo {year} {2025})},\ \Eprint
  {https://arxiv.org/abs/2406.04894} {arXiv:2406.04894 [astro-ph.CO]}
  \BibitemShut {NoStop}%
\bibitem [{\citenamefont {{Criswell}}\ \emph
  {et~al.}(2025{\natexlab{b}})\citenamefont {{Criswell}}, \citenamefont
  {{Banagiri}}, \citenamefont {{Lawrence}}, \citenamefont {{Schult}},
  \citenamefont {{Rieck}}, \citenamefont {{Taylor}},\ and\ \citenamefont
  {{Mandic}}}]{2025arXiv250820308C}%
  \BibitemOpen
  \bibfield  {author} {\bibinfo {author} {\bibfnamefont {A.~W.}\ \bibnamefont
  {{Criswell}}}, \bibinfo {author} {\bibfnamefont {S.}~\bibnamefont
  {{Banagiri}}}, \bibinfo {author} {\bibfnamefont {J.}~\bibnamefont
  {{Lawrence}}}, \bibinfo {author} {\bibfnamefont {L.}~\bibnamefont
  {{Schult}}}, \bibinfo {author} {\bibfnamefont {S.}~\bibnamefont {{Rieck}}},
  \bibinfo {author} {\bibfnamefont {S.~R.}\ \bibnamefont {{Taylor}}},\ and\
  \bibinfo {author} {\bibfnamefont {V.}~\bibnamefont {{Mandic}}},\ }\href
  {https://doi.org/10.48550/arXiv.2508.20308} {\bibfield  {journal} {\bibinfo
  {journal} {arXiv e-prints}\ ,\ \bibinfo {eid} {arXiv:2508.20308}} (\bibinfo
  {year} {2025}{\natexlab{b}})},\ \Eprint {https://arxiv.org/abs/2508.20308}
  {arXiv:2508.20308 [astro-ph.IM]} \BibitemShut {NoStop}%
\bibitem [{\citenamefont {{Banagiri}}\ \emph {et~al.}(2021)\citenamefont
  {{Banagiri}}, \citenamefont {{Criswell}}, \citenamefont {{Kuan}},
  \citenamefont {{Mandic}}, \citenamefont {{Romano}},\ and\ \citenamefont
  {{Taylor}}}]{2021MNRAS.507.5451B}%
  \BibitemOpen
  \bibfield  {author} {\bibinfo {author} {\bibfnamefont {S.}~\bibnamefont
  {{Banagiri}}}, \bibinfo {author} {\bibfnamefont {A.}~\bibnamefont
  {{Criswell}}}, \bibinfo {author} {\bibfnamefont {T.}~\bibnamefont {{Kuan}}},
  \bibinfo {author} {\bibfnamefont {V.}~\bibnamefont {{Mandic}}}, \bibinfo
  {author} {\bibfnamefont {J.~D.}\ \bibnamefont {{Romano}}},\ and\ \bibinfo
  {author} {\bibfnamefont {S.~R.}\ \bibnamefont {{Taylor}}},\ }\href
  {https://doi.org/10.1093/mnras/stab2479} {\bibfield  {journal} {\bibinfo
  {journal} {\mnras}\ }\textbf {\bibinfo {volume} {507}},\ \bibinfo {pages}
  {5451} (\bibinfo {year} {2021})},\ \Eprint {https://arxiv.org/abs/2103.00826}
  {arXiv:2103.00826 [astro-ph.IM]} \BibitemShut {NoStop}%
\bibitem [{LDC()}]{LDC1b}%
  \BibitemOpen
  \href@noop {} {\bibinfo {title} {Lisa data challenge 1b: Yorsh}},\ \bibinfo
  {howpublished} {\url{https://sbgvm-151-90.in2p3.fr/challenge1b}}\BibitemShut
  {NoStop}%
\bibitem [{\citenamefont {{Pozzoli}}\ \emph
  {et~al.}(2025{\natexlab{b}})\citenamefont {{Pozzoli}}, \citenamefont
  {{Buscicchio}}, \citenamefont {{Klein}},\ and\ \citenamefont
  {{Chirico}}}]{2025arXiv250622542P}%
  \BibitemOpen
  \bibfield  {author} {\bibinfo {author} {\bibfnamefont {F.}~\bibnamefont
  {{Pozzoli}}}, \bibinfo {author} {\bibfnamefont {R.}~\bibnamefont
  {{Buscicchio}}}, \bibinfo {author} {\bibfnamefont {A.}~\bibnamefont
  {{Klein}}},\ and\ \bibinfo {author} {\bibfnamefont {D.}~\bibnamefont
  {{Chirico}}},\ }\href {https://doi.org/10.48550/arXiv.2506.22542} {\bibfield
  {journal} {\bibinfo  {journal} {arXiv e-prints}\ ,\ \bibinfo {eid}
  {arXiv:2506.22542}} (\bibinfo {year} {2025}{\natexlab{b}})},\ \Eprint
  {https://arxiv.org/abs/2506.22542} {arXiv:2506.22542 [astro-ph.IM]}
  \BibitemShut {NoStop}%
\bibitem [{\citenamefont {{Tinto}}\ and\ \citenamefont
  {{Dhurandhar}}(2005)}]{Tinto:2005}%
  \BibitemOpen
  \bibfield  {author} {\bibinfo {author} {\bibfnamefont {M.}~\bibnamefont
  {{Tinto}}}\ and\ \bibinfo {author} {\bibfnamefont {S.~V.}\ \bibnamefont
  {{Dhurandhar}}},\ }\href {https://doi.org/10.12942/lrr-2005-4} {\bibfield
  {journal} {\bibinfo  {journal} {Living Reviews in Relativity}\ }\textbf
  {\bibinfo {volume} {8}},\ \bibinfo {eid} {4} (\bibinfo {year}
  {2005})}\BibitemShut {NoStop}%
\bibitem [{\citenamefont {{Kume}}\ \emph {et~al.}(2025)\citenamefont {{Kume}},
  \citenamefont {{Peloso}}, \citenamefont {{Pieroni}},\ and\ \citenamefont
  {{Ricciardone}}}]{2025JCAP...06..030K}%
  \BibitemOpen
  \bibfield  {author} {\bibinfo {author} {\bibfnamefont {J.}~\bibnamefont
  {{Kume}}}, \bibinfo {author} {\bibfnamefont {M.}~\bibnamefont {{Peloso}}},
  \bibinfo {author} {\bibfnamefont {M.}~\bibnamefont {{Pieroni}}},\ and\
  \bibinfo {author} {\bibfnamefont {A.}~\bibnamefont {{Ricciardone}}},\ }\href
  {https://doi.org/10.1088/1475-7516/2025/06/030} {\bibfield  {journal}
  {\bibinfo  {journal} {\jcap}\ }\textbf {\bibinfo {volume} {2025}},\ \bibinfo
  {eid} {030} (\bibinfo {year} {2025})},\ \Eprint
  {https://arxiv.org/abs/2410.10342} {arXiv:2410.10342 [gr-qc]} \BibitemShut
  {NoStop}%
\bibitem [{\citenamefont {Welch}(1967)}]{1161901}%
  \BibitemOpen
  \bibfield  {author} {\bibinfo {author} {\bibfnamefont {P.}~\bibnamefont
  {Welch}},\ }\href {https://doi.org/10.1109/TAU.1967.1161901} {\bibfield
  {journal} {\bibinfo  {journal} {IEEE Transactions on Audio and
  Electroacoustics}\ }\textbf {\bibinfo {volume} {15}},\ \bibinfo {pages} {70}
  (\bibinfo {year} {1967})}\BibitemShut {NoStop}%
\bibitem [{\citenamefont {{Karnesis}}\ \emph {et~al.}(2021)\citenamefont
  {{Karnesis}}, \citenamefont {{Babak}}, \citenamefont {{Pieroni}},
  \citenamefont {{Cornish}},\ and\ \citenamefont
  {{Littenberg}}}]{Karnesis:2021}%
  \BibitemOpen
  \bibfield  {author} {\bibinfo {author} {\bibfnamefont {N.}~\bibnamefont
  {{Karnesis}}}, \bibinfo {author} {\bibfnamefont {S.}~\bibnamefont {{Babak}}},
  \bibinfo {author} {\bibfnamefont {M.}~\bibnamefont {{Pieroni}}}, \bibinfo
  {author} {\bibfnamefont {N.}~\bibnamefont {{Cornish}}},\ and\ \bibinfo
  {author} {\bibfnamefont {T.}~\bibnamefont {{Littenberg}}},\ }\href
  {https://doi.org/10.1103/PhysRevD.104.043019} {\bibfield  {journal} {\bibinfo
   {journal} {\prd}\ }\textbf {\bibinfo {volume} {104}},\ \bibinfo {eid}
  {043019} (\bibinfo {year} {2021})},\ \Eprint
  {https://arxiv.org/abs/2103.14598} {arXiv:2103.14598 [astro-ph.IM]}
  \BibitemShut {NoStop}%
\bibitem [{\citenamefont {{Baghi}}\ \emph {et~al.}(2023)\citenamefont
  {{Baghi}}, \citenamefont {{Karnesis}}, \citenamefont {{Bayle}}, \citenamefont
  {{Besan{\c{c}}on}},\ and\ \citenamefont
  {{Inchausp{\'e}}}}]{2023JCAP...04..066B}%
  \BibitemOpen
  \bibfield  {author} {\bibinfo {author} {\bibfnamefont {Q.}~\bibnamefont
  {{Baghi}}}, \bibinfo {author} {\bibfnamefont {N.}~\bibnamefont {{Karnesis}}},
  \bibinfo {author} {\bibfnamefont {J.-B.}\ \bibnamefont {{Bayle}}}, \bibinfo
  {author} {\bibfnamefont {M.}~\bibnamefont {{Besan{\c{c}}on}}},\ and\ \bibinfo
  {author} {\bibfnamefont {H.}~\bibnamefont {{Inchausp{\'e}}}},\ }\href
  {https://doi.org/10.1088/1475-7516/2023/04/066} {\bibfield  {journal}
  {\bibinfo  {journal} {\jcap}\ }\textbf {\bibinfo {volume} {2023}},\ \bibinfo
  {eid} {066} (\bibinfo {year} {2023})},\ \Eprint
  {https://arxiv.org/abs/2302.12573} {arXiv:2302.12573 [gr-qc]} \BibitemShut
  {NoStop}%
\bibitem [{\citenamefont {{Pozzoli}}\ \emph {et~al.}(2024)\citenamefont
  {{Pozzoli}}, \citenamefont {{Buscicchio}}, \citenamefont {{Moore}},
  \citenamefont {{Haardt}},\ and\ \citenamefont
  {{Sesana}}}]{2024PhRvD.109h3029P}%
  \BibitemOpen
  \bibfield  {author} {\bibinfo {author} {\bibfnamefont {F.}~\bibnamefont
  {{Pozzoli}}}, \bibinfo {author} {\bibfnamefont {R.}~\bibnamefont
  {{Buscicchio}}}, \bibinfo {author} {\bibfnamefont {C.~J.}\ \bibnamefont
  {{Moore}}}, \bibinfo {author} {\bibfnamefont {F.}~\bibnamefont {{Haardt}}},\
  and\ \bibinfo {author} {\bibfnamefont {A.}~\bibnamefont {{Sesana}}},\ }\href
  {https://doi.org/10.1103/PhysRevD.109.083029} {\bibfield  {journal} {\bibinfo
   {journal} {\prd}\ }\textbf {\bibinfo {volume} {109}},\ \bibinfo {eid}
  {083029} (\bibinfo {year} {2024})},\ \Eprint
  {https://arxiv.org/abs/2311.12111} {arXiv:2311.12111 [astro-ph.CO]}
  \BibitemShut {NoStop}%
\bibitem [{\citenamefont {{Quang Nam}}\ \emph {et~al.}(2023)\citenamefont
  {{Quang Nam}}, \citenamefont {{Martino}}, \citenamefont {{Lemi{\`e}re}},
  \citenamefont {{Petiteau}}, \citenamefont {{Bayle}}, \citenamefont
  {{Hartwig}},\ and\ \citenamefont {{Staab}}}]{Nam:2023}%
  \BibitemOpen
  \bibfield  {author} {\bibinfo {author} {\bibfnamefont {D.}~\bibnamefont
  {{Quang Nam}}}, \bibinfo {author} {\bibfnamefont {J.}~\bibnamefont
  {{Martino}}}, \bibinfo {author} {\bibfnamefont {Y.}~\bibnamefont
  {{Lemi{\`e}re}}}, \bibinfo {author} {\bibfnamefont {A.}~\bibnamefont
  {{Petiteau}}}, \bibinfo {author} {\bibfnamefont {J.-B.}\ \bibnamefont
  {{Bayle}}}, \bibinfo {author} {\bibfnamefont {O.}~\bibnamefont {{Hartwig}}},\
  and\ \bibinfo {author} {\bibfnamefont {M.}~\bibnamefont {{Staab}}},\ }\href
  {https://doi.org/10.1103/PhysRevD.108.082004} {\bibfield  {journal} {\bibinfo
   {journal} {\prd}\ }\textbf {\bibinfo {volume} {108}},\ \bibinfo {eid}
  {082004} (\bibinfo {year} {2023})},\ \Eprint
  {https://arxiv.org/abs/2211.02539} {arXiv:2211.02539 [gr-qc]} \BibitemShut
  {NoStop}%
\bibitem [{\citenamefont {{Williams}}\ \emph {et~al.}(2021)\citenamefont
  {{Williams}}, \citenamefont {{Veitch}},\ and\ \citenamefont
  {{Messenger}}}]{2021PhRvD.103j3006W}%
  \BibitemOpen
  \bibfield  {author} {\bibinfo {author} {\bibfnamefont {M.~J.}\ \bibnamefont
  {{Williams}}}, \bibinfo {author} {\bibfnamefont {J.}~\bibnamefont
  {{Veitch}}},\ and\ \bibinfo {author} {\bibfnamefont {C.}~\bibnamefont
  {{Messenger}}},\ }\href {https://doi.org/10.1103/PhysRevD.103.103006}
  {\bibfield  {journal} {\bibinfo  {journal} {\prd}\ }\textbf {\bibinfo
  {volume} {103}},\ \bibinfo {eid} {103006} (\bibinfo {year} {2021})},\ \Eprint
  {https://arxiv.org/abs/2102.11056} {arXiv:2102.11056 [gr-qc]} \BibitemShut
  {NoStop}%
\bibitem [{\citenamefont {{Phan}}\ \emph {et~al.}(2019)\citenamefont {{Phan}},
  \citenamefont {{Pradhan}},\ and\ \citenamefont
  {{Jankowiak}}}]{2019arXiv191211554P}%
  \BibitemOpen
  \bibfield  {author} {\bibinfo {author} {\bibfnamefont {D.}~\bibnamefont
  {{Phan}}}, \bibinfo {author} {\bibfnamefont {N.}~\bibnamefont {{Pradhan}}},\
  and\ \bibinfo {author} {\bibfnamefont {M.}~\bibnamefont {{Jankowiak}}},\
  }\href {https://doi.org/10.48550/arXiv.1912.11554} {\bibfield  {journal}
  {\bibinfo  {journal} {arXiv e-prints}\ ,\ \bibinfo {eid} {arXiv:1912.11554}}
  (\bibinfo {year} {2019})},\ \Eprint {https://arxiv.org/abs/1912.11554}
  {arXiv:1912.11554 [stat.ML]} \BibitemShut {NoStop}%
\bibitem [{\citenamefont {{Franciolini}}\ \emph {et~al.}(2025)\citenamefont
  {{Franciolini}}, \citenamefont {{Pieroni}}, \citenamefont {{Ricciardone}},\
  and\ \citenamefont {{Romano}}}]{2025arXiv250524695F}%
  \BibitemOpen
  \bibfield  {author} {\bibinfo {author} {\bibfnamefont {G.}~\bibnamefont
  {{Franciolini}}}, \bibinfo {author} {\bibfnamefont {M.}~\bibnamefont
  {{Pieroni}}}, \bibinfo {author} {\bibfnamefont {A.}~\bibnamefont
  {{Ricciardone}}},\ and\ \bibinfo {author} {\bibfnamefont {J.~D.}\
  \bibnamefont {{Romano}}},\ }\href {https://doi.org/10.48550/arXiv.2505.24695}
  {\bibfield  {journal} {\bibinfo  {journal} {arXiv e-prints}\ ,\ \bibinfo
  {eid} {arXiv:2505.24695}} (\bibinfo {year} {2025})},\ \Eprint
  {https://arxiv.org/abs/2505.24695} {arXiv:2505.24695 [gr-qc]} \BibitemShut
  {NoStop}%
\bibitem [{\citenamefont {Blackwell}(1947)}]{Blackwell1947}%
  \BibitemOpen
  \bibfield  {author} {\bibinfo {author} {\bibfnamefont {D.}~\bibnamefont
  {Blackwell}},\ }\href {http://www.jstor.org/stable/2236107} {\bibfield
  {journal} {\bibinfo  {journal} {The Annals of Mathematical Statistics}\
  }\textbf {\bibinfo {volume} {18}},\ \bibinfo {pages} {105} (\bibinfo {year}
  {1947})}\BibitemShut {NoStop}%
\bibitem [{\citenamefont {Rao}(1992)}]{Rao1992}%
  \BibitemOpen
  \bibfield  {author} {\bibinfo {author} {\bibfnamefont {C.~R.}\ \bibnamefont
  {Rao}},\ }\bibinfo {title} {Information and the accuracy attainable in the
  estimation of statistical parameters},\ in\ \href
  {https://doi.org/10.1007/978-1-4612-0919-5_16} {\emph {\bibinfo {booktitle}
  {Breakthroughs in Statistics: Foundations and Basic Theory}}},\ \bibinfo
  {editor} {edited by\ \bibinfo {editor} {\bibfnamefont {S.}~\bibnamefont
  {Kotz}}\ and\ \bibinfo {editor} {\bibfnamefont {N.~L.}\ \bibnamefont
  {Johnson}}}\ (\bibinfo  {publisher} {Springer New York},\ \bibinfo {address}
  {New York, NY},\ \bibinfo {year} {1992})\ pp.\ \bibinfo {pages}
  {235--247}\BibitemShut {NoStop}%
\bibitem [{\citenamefont {Lin}(1991)}]{61115}%
  \BibitemOpen
  \bibfield  {author} {\bibinfo {author} {\bibfnamefont {J.}~\bibnamefont
  {Lin}},\ }\href {https://doi.org/10.1109/18.61115} {\bibfield  {journal}
  {\bibinfo  {journal} {IEEE Transactions on Information Theory}\ }\textbf
  {\bibinfo {volume} {37}},\ \bibinfo {pages} {145} (\bibinfo {year}
  {1991})}\BibitemShut {NoStop}%
\bibitem [{\citenamefont {Skilling}(2006)}]{10.1214/06-BA127}%
  \BibitemOpen
  \bibfield  {author} {\bibinfo {author} {\bibfnamefont {J.}~\bibnamefont
  {Skilling}},\ }\href {https://doi.org/10.1214/06-BA127} {\bibfield  {journal}
  {\bibinfo  {journal} {Bayesian Analysis}\ }\textbf {\bibinfo {volume} {1}},\
  \bibinfo {pages} {833 } (\bibinfo {year} {2006})}\BibitemShut {NoStop}%
\bibitem [{\citenamefont {Maturana-Russel}\ \emph {et~al.}(2019)\citenamefont
  {Maturana-Russel}, \citenamefont {Meyer}, \citenamefont {Veitch},\ and\
  \citenamefont {Christensen}}]{Maturana:2019}%
  \BibitemOpen
  \bibfield  {author} {\bibinfo {author} {\bibfnamefont {P.}~\bibnamefont
  {Maturana-Russel}}, \bibinfo {author} {\bibfnamefont {R.}~\bibnamefont
  {Meyer}}, \bibinfo {author} {\bibfnamefont {J.}~\bibnamefont {Veitch}},\ and\
  \bibinfo {author} {\bibfnamefont {N.}~\bibnamefont {Christensen}},\ }\href
  {https://doi.org/10.1103/PhysRevD.99.084006} {\bibfield  {journal} {\bibinfo
  {journal} {Physical Review D}\ }\textbf {\bibinfo {volume} {99}},\ \bibinfo
  {pages} {084006} (\bibinfo {year} {2019})}\BibitemShut {NoStop}%
\bibitem [{\citenamefont {{Zahraoui}}\ \emph {et~al.}(2025)\citenamefont
  {{Zahraoui}}, \citenamefont {{Maturana-Russel}}, \citenamefont {{van
  Straten}}, \citenamefont {{Meyer}},\ and\ \citenamefont
  {{Gulyaev}}}]{2025MNRAS.540.3818Z}%
  \BibitemOpen
  \bibfield  {author} {\bibinfo {author} {\bibfnamefont {E.~M.}\ \bibnamefont
  {{Zahraoui}}}, \bibinfo {author} {\bibfnamefont {P.}~\bibnamefont
  {{Maturana-Russel}}}, \bibinfo {author} {\bibfnamefont {W.}~\bibnamefont
  {{van Straten}}}, \bibinfo {author} {\bibfnamefont {R.}~\bibnamefont
  {{Meyer}}},\ and\ \bibinfo {author} {\bibfnamefont {S.}~\bibnamefont
  {{Gulyaev}}},\ }\href {https://doi.org/10.1093/mnras/staf953} {\bibfield
  {journal} {\bibinfo  {journal} {\mnras}\ }\textbf {\bibinfo {volume} {540}},\
  \bibinfo {pages} {3818} (\bibinfo {year} {2025})},\ \Eprint
  {https://arxiv.org/abs/2411.14736} {arXiv:2411.14736 [astro-ph.IM]}
  \BibitemShut {NoStop}%
\bibitem [{\citenamefont {{Burke}}\ \emph {et~al.}(2025)\citenamefont
  {{Burke}}, \citenamefont {{Marsat}}, \citenamefont {{Gair}},\ and\
  \citenamefont {{Katz}}}]{2025PhRvD.111l4053B}%
  \BibitemOpen
  \bibfield  {author} {\bibinfo {author} {\bibfnamefont {O.}~\bibnamefont
  {{Burke}}}, \bibinfo {author} {\bibfnamefont {S.}~\bibnamefont {{Marsat}}},
  \bibinfo {author} {\bibfnamefont {J.~R.}\ \bibnamefont {{Gair}}},\ and\
  \bibinfo {author} {\bibfnamefont {M.~L.}\ \bibnamefont {{Katz}}},\ }\href
  {https://doi.org/10.1103/5jr8-k2ss} {\bibfield  {journal} {\bibinfo
  {journal} {\prd}\ }\textbf {\bibinfo {volume} {111}},\ \bibinfo {eid}
  {124053} (\bibinfo {year} {2025})},\ \Eprint
  {https://arxiv.org/abs/2502.17426} {arXiv:2502.17426 [gr-qc]} \BibitemShut
  {NoStop}%
\bibitem [{\citenamefont {{Dey}}\ \emph {et~al.}(2021)\citenamefont {{Dey}},
  \citenamefont {{Karnesis}}, \citenamefont {{Toubiana}}, \citenamefont
  {{Barausse}}, \citenamefont {{Korsakova}}, \citenamefont {{Baghi}},\ and\
  \citenamefont {{Basak}}}]{2021PhRvD.104d4035D}%
  \BibitemOpen
  \bibfield  {author} {\bibinfo {author} {\bibfnamefont {K.}~\bibnamefont
  {{Dey}}}, \bibinfo {author} {\bibfnamefont {N.}~\bibnamefont {{Karnesis}}},
  \bibinfo {author} {\bibfnamefont {A.}~\bibnamefont {{Toubiana}}}, \bibinfo
  {author} {\bibfnamefont {E.}~\bibnamefont {{Barausse}}}, \bibinfo {author}
  {\bibfnamefont {N.}~\bibnamefont {{Korsakova}}}, \bibinfo {author}
  {\bibfnamefont {Q.}~\bibnamefont {{Baghi}}},\ and\ \bibinfo {author}
  {\bibfnamefont {S.}~\bibnamefont {{Basak}}},\ }\href
  {https://doi.org/10.1103/PhysRevD.104.044035} {\bibfield  {journal} {\bibinfo
   {journal} {\prd}\ }\textbf {\bibinfo {volume} {104}},\ \bibinfo {eid}
  {044035} (\bibinfo {year} {2021})},\ \Eprint
  {https://arxiv.org/abs/2104.12646} {arXiv:2104.12646 [gr-qc]} \BibitemShut
  {NoStop}%
\bibitem [{\citenamefont {Buscicchio}\ \emph {et~al.}(2025)\citenamefont
  {Buscicchio}, \citenamefont {Pozzoli}, \citenamefont {Chirico},\ and\
  \citenamefont {Sesana}}]{bahamasgf-release}%
  \BibitemOpen
  \bibfield  {author} {\bibinfo {author} {\bibfnamefont {R.}~\bibnamefont
  {Buscicchio}}, \bibinfo {author} {\bibfnamefont {F.}~\bibnamefont {Pozzoli}},
  \bibinfo {author} {\bibfnamefont {D.}~\bibnamefont {Chirico}},\ and\ \bibinfo
  {author} {\bibfnamefont {A.}~\bibnamefont {Sesana}},\ }\href
  {https://doi.org/10.5281/zenodo.17527235} {\bibinfo {title}
  {Riccardobuscicchio/bahamas-galactic-foreground- release: Release for paper
  submission}} (\bibinfo {year} {2025})\BibitemShut {NoStop}%
\end{thebibliography}%

\onecolumngrid
\clearpage

\appendix
\renewcommand{\thefigure}{A\arabic{figure}}
\setcounter{figure}{0} 

\section{Corner and ridge posterior plots}
\label{app:plot}

\begin{figure*}[htbp]
    \centering
    \includegraphics[width=0.95\columnwidth]{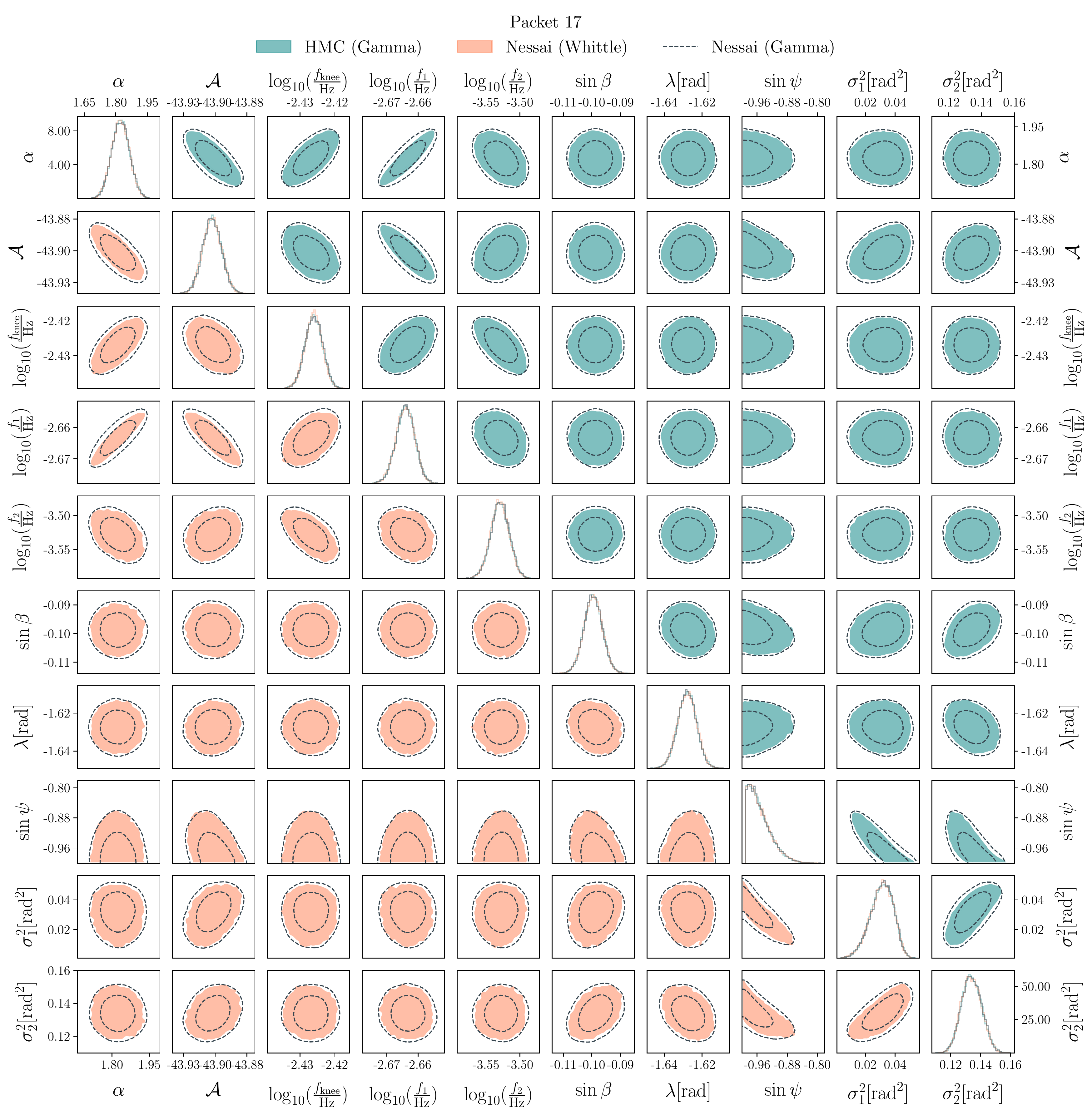}
    \caption{Posterior distributions of the parameters obtained with different \texttt{bahamas} setup configurations. The upper right triangle plot compares the results from NS (dashed contour line) and NUTS (teal filled contour) sampling, both using the Gamma likelihood.
    The lower left triangle plot shows a comparison between two likelihood choices, Whittle (coral filled contours) and Gamma (dashed contour line), both employing NS algorithm. Overall, no significant differences appear in the posterior distributions, demonstrating consistency across all the proposed setups. The posterior samples are obtained from the sequential analysis up to packet 17, corresponding to 34 consecutive weeks of data.}
    \label{fig:corner_sum}
\end{figure*}

\begin{figure*}[t]
    \centering
    \includegraphics[width=1\columnwidth]{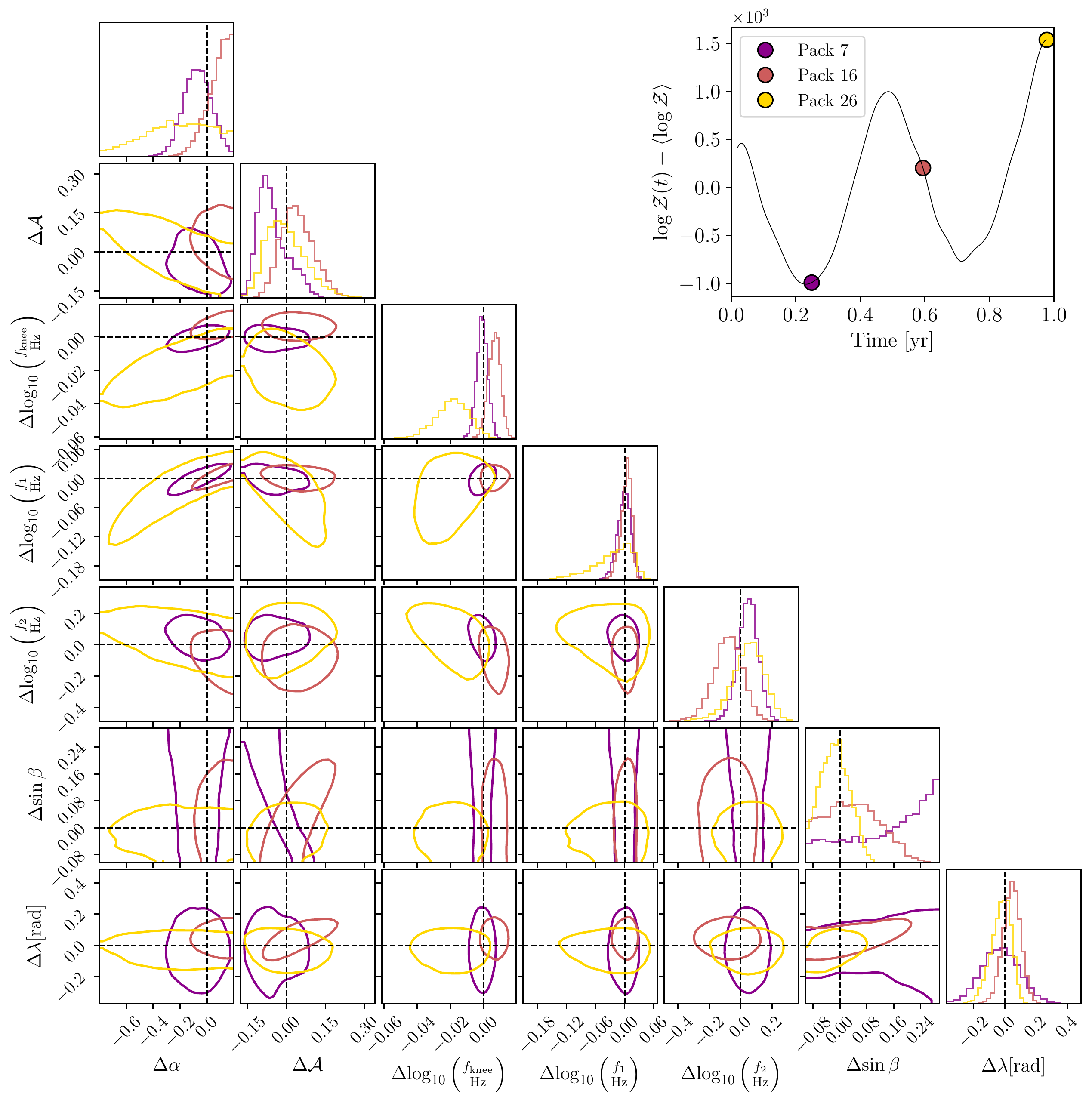}
    \caption{Posterior distributions of the spectral parameters of the Galactic foreground and the sky coordinates of the Gaussian center describing its spatial distribution. Posteriors contours are shown relative to their true values. The color-coded posteriors correspond to three different values of the evidence for the quasi-stationary model in the differential analysis with respect to the average over the whole first year of LISA data. In particular, we consider three values (top-right inset) below, above, and comparable to the average, in purple, yellow, and coral, respectively. 
    The plot illustrates how a reconstruction of the sky position is associated to higher evidences, while the precision on the spectral parameters degrades. 
    }
    \label{fig:corner_chunk}
\end{figure*}

\begin{figure*}
    \centering
\includegraphics[width=1\columnwidth]{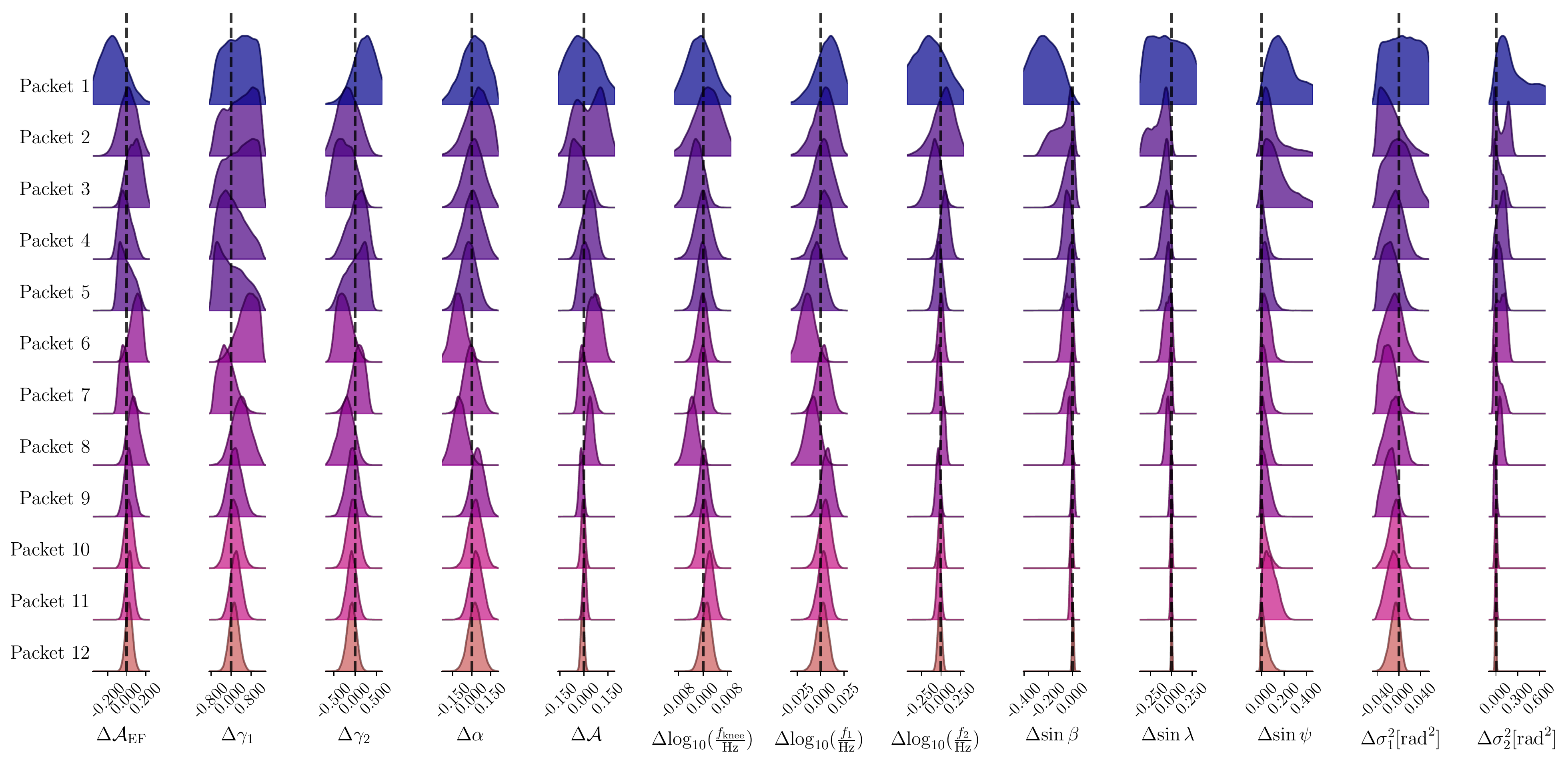}
    \includegraphics[width=1\columnwidth]{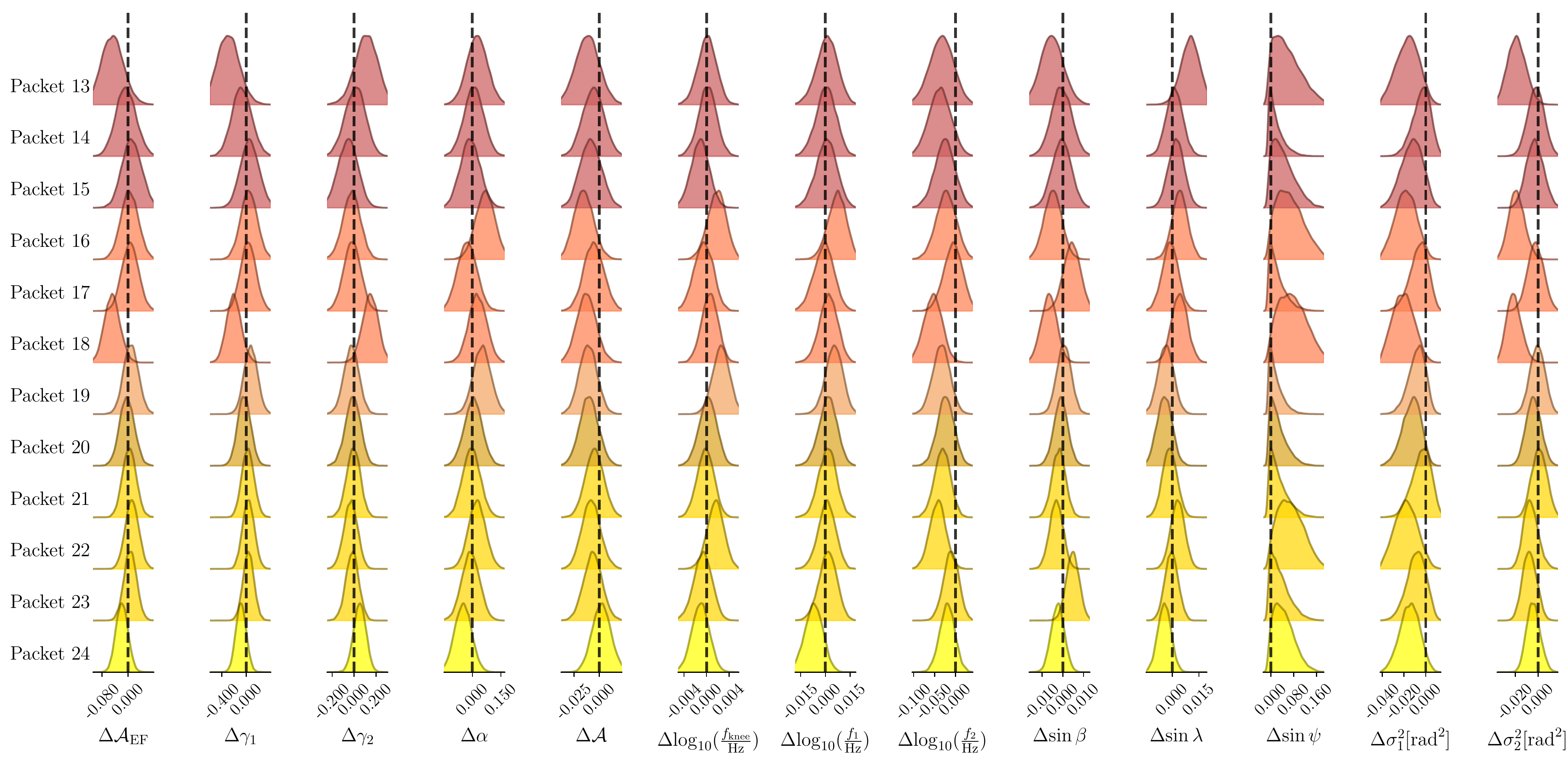}
    \caption{Marginal posterior distributions on extragalactic and Galactic foreground parameters from a sequential, one-year, two-weeks-segments analysis. 
    Posteriors are shifted to the injected values, indicating unbiased reconstruction. Individual posterior colors follow the same convention as~\Cref{fig:ridgeline}. The extragalactic foreground becomes well constrained starting from Packet 7, corresponding to roughly 3 months of observed data.}
    \label{fig:ridgeline2}
\end{figure*}
\twocolumngrid
\end{document}